\documentclass[review]{elsarticle}

\usepackage{lineno,hyperref}
\modulolinenumbers[5]

\usepackage[letterpaper]{geometry}
\usepackage{amsmath}
\usepackage{amsfonts}
\usepackage{relsize}
\usepackage{braket}

\newcommand{\negtab}{\hspace*{-2em}}

\newcommand{\setsd}{\ensuremath S_{\mathbf{m}}}
\newcommand{\setexc}{\ensuremath S_{\hspace{-0.1em}\hat{E}}}

\makeatletter
\let\ams@underbrace=\underbrace
\def\underbrace#1_#2{%
  \setbox0=\hbox{$\displaystyle#1$}%
  \ams@underbrace{#1}_{\parbox[t]{\the\wd0}{\centering#2}}%
}
\makeatother

\makeatletter
\def\@author#1{\g@addto@macro\elsauthors{\normalsize%
    \def\baselinestretch{1}%
    \upshape\authorsep#1\unskip\textsuperscript{%
      \ifx\@fnmark\@empty\else\unskip\sep\@fnmark\let\sep=,\fi
      \ifx\@corref\@empty\else\unskip\sep\@corref\let\sep=,\fi
      }%
    \def\authorsep{\unskip,\space}%
    \global\let\@fnmark\@empty
    \global\let\@corref\@empty  %% Added
    \global\let\sep\@empty}%
    \@eadauthor={#1}
}
\makeatother

\journal{Computational and Theoretical Chemistry}
\begin{document}

\begin{frontmatter}

\title{Flexible Ansatz for N-body Configuration Interaction}

%% Group authors per affiliation:
\author[a]{Taewon D. Kim}
\author[b]{Ram\'{o}n Alain Miranda-Quintana}
\author[a]{Michael Richer}
\author[a]{Paul W. Ayers\corref{c}}
\address[a]{Department of Chemistry and Chemical Biology, McMaster University, Hamilton, Ontario, Canada L8S 4M1}
\address[b]{Department of Chemistry, University of Florida, Gainesville, FL, 32603 USA}
\cortext[c]{Corresponding Author; ayers@mcmaster.ca}

\begin{abstract}
We present a Flexible Ansatz for N-body Configuration Interaction
(FANCI) that includes any multideterminant wavefunction.
This ansatz is a generalization of the Configuration Interaction (CI)
wavefunction, where the coefficients are replaced by a specified function of
certain parameters.
By making an appropriate choice for this function, we can reproduce popular
wavefunction structures like CI, Coupled-Cluster, Tensor Network States, and
geminal-product wavefunctions.
The universality of this framework suggests a programming structure that
allows for the easy construction and optimization of arbitrary wavefunctions.
Here, we will discuss the structures of the FANCI framework and its implications
for wavefunction properties, particularly accuracy, cost, and
size-consistency.
We demonstrate the flexibility of this framework by reconstructing popular
wavefunction ans\"{a}tze and modifying them to construct novel wavefunction forms.
FANCI provides a powerful framework for exploring, developing, and testing new
wavefunction forms.
\end{abstract}


\end{frontmatter}

\linenumbers

% \newpage
% \tableofcontents
% \newpage

\section{Introduction}
\label{sec:intro}
In this paper, we focus on electronic systems, whose Hamiltonian can be written
as
\begin{equation}
  \hat{H}_{elec} = \sum_{ij} h_{ij} a^\dagger_i a_j
  + \frac{1}{2} \sum_{ijkl} g_{ijkl} a^\dagger_i a^\dagger_j a_l a_k
\end{equation}
where $h_{ij}$ and $g_{ijkl}$ are the one- and two-electron integrals and
$a^\dagger_i$ ($a_i$) creates (annihilates) the $i$\textsuperscript{th}
spin-orbital.
The exact solutions to the electronic Hamiltonian can be written as a linear
combination of all possible $N$-electron basis functions (Slater determinants)
formed from the given set of spin-orbitals.
This is the Full Configuration Interaction (FCI) wavefunction\cite{fci}:
\begin{equation}
  \label{eq:fci}
  \ket{\Psi_{\mathrm{FCI}}} = \sum_{\mathbf{m}}^{\binom{2K}{N}} C_{\mathbf{m}} \ket{\mathbf{m}}
\end{equation}
where $2K$ is the number of spin-orbitals, $N$ is the number of electrons, and
$C_m$ is the coefficient of the Slater determinant $\ket{\mathbf{m}}$.
We can think of $\mathbf{m}$ as an occupation vector that specifies which of the
$2K$ spin-orbitals are occupied to construct the $N$-electron basis functions.
The number of parameters for the FCI wavefunction scales combinatorially
with the number of orbitals and electrons, so brute-force direct
calculations of the FCI wavefunctions are restricted to small systems with small
basis sets.

Various approximations can be made to the Schr\"{o}dinger equation to bring down
its cost:
(1) simplify the Hamiltonian,
(2) find alternative algorithms,
and (3) parameterize the FCI wavefunction\cite{1, 2, 3, 4, 5, 6, proj_var,
  dmrg_intro, proj_schr_3}.
In this article, we focus on the parameterization of the FCI wavefunction.
The simplest approximation to the FCI wavefunction involves explicitly selecting
(or truncating) the Slater determinants that contribute to the wavefunction.
Such wavefunctions are broadly termed selected Configuration Interaction (CI)
wavefunctions:
\begin{equation}
  \label{eq:ci}
  \begin{split}
    \ket{\Psi_{\mathrm{CI}}}
    &= \sum_{\mathbf{m} \in S} C_{\mathbf{m}} \ket{\mathbf{m}}\\
  \end{split}
\end{equation}
where $S$ is a subset of the Slater determinants within the given basis.
The Slater determinants can be selected by seniority, such as the doubly occupied CI
(DOCI) wavefunction\cite{7, 8, 9, 10, 11, 12, 13, 14, 15, 16, 17, doci},
or by excitation-level relative to a reference Slater determinant, such as CI
singles and doubles (CISD) wavefunction\cite{cisd}.
Alternatively, we can select all (or many) of the Slater determinants in a given
set of orbitals.
This leads to active-space methods like CASSCF\cite{casscf_?}, RASSCF
\cite{rasscf}, and MCSCF\cite{mcscf}.
Finally, if the orbitals are localized, the Slater determinants that embody chemically
intuitive concepts can be linearly combined to construct Valence Bond (VB)
structures\cite{18, 19, 20, 21, vb}.
This leads to VBCI methods.
While there are many variants of the CI wavefunction, most are not
size-consistent and choosing an efficient set of orbitals/determinants is
molecule and geometry dependent.
For truly strongly-correlated systems, which have myriad Slater determinants
with small yet significant contributions, selected CI methods generally fail.

Alternatively, the FCI wavefunction can be approximated by an alternative form
(ansatz), such as a nonlinear function of parameters that are not simply the
coefficients of the Slater determinants.
For example, the Coupled-Cluster (CC) wavefunction parameterizes the CI wavefunction using an
exponential ansatz\cite{22, 23, 24, cc_review, ramonemail_cc1, ramonemail_cc2}:
\begin{equation}
  \label{eq:cc}
  \begin{split}
    \ket{\Psi_{\mathrm{CC}}}
    &= \exp\left( \sum_{i} \sum_a t_i^a \hat{E}_i^a
      + \sum_{i<j} \sum_{a<b} t_{ij}^{ab} \hat{E}_{ij}^{ab}
      + \dots \right) \ket{\Phi_{\mathrm{HF}}}\\
    &= \exp\left(
      \sum_{
        \begin{smallmatrix}
          \mathbf{i}, \mathbf{a}\\
          \hat{E}_{\mathbf{i}}^{\mathbf{a}} \in \tilde{S}_{\hat{E}}
        \end{smallmatrix}
      }
      t_{\mathbf{i}}^{\mathbf{a}} \hat{E}_{\mathbf{i}}^{\mathbf{a}}
    \right)\ket{\Phi_{\mathrm{HF}}}\\
  \end{split}
\end{equation}
where $\hat{E}_{\mathbf{i}}^{\mathbf{a}}$ is an excitation operator that excites
electrons from a set of occupied orbitals $\mathbf{i}$ to a set of virtual
orbitals $\mathbf{a}$, $t_{\mathbf{i}}^{\mathbf{a}}$ is the associated
coefficient/amplitude, and $\tilde{S}_{\hat{E}}$ is the set of allowed excitation operators.
Just as in CI wavefunctions, the complexity of the wavefunction (and the number
of parameters) can be controlled by truncating the set of excitation operators
used.

Tensor Product State (TPS) wavefunctions are expressed with respect to
parameters that describe the correlations between spatial orbitals of different
states (occupations)\cite{25, 26, 27, 28, tns_intro, paulemail1, paulemail2, paulemail3, paulemail4, paulemail5, paulemail6, paulemail7, paulemail8}:
\begin{equation}
  \label{eq:tps}
  \begin{split}
    \ket{\Psi_{\mathrm{TPS}}} &= \sum_{
      \begin{smallmatrix}
        n_1 \dots n_K
      \end{smallmatrix}
    } \sum_{
      \begin{smallmatrix}
        i_{12} \dots i_{1K}\\
        i_{23} \dots i_{2K}\\
        \vdots\\
        i_{K-1\, K}
      \end{smallmatrix}
    } (M_1)_{i_{12} \dots i_{1K}}^{n_1}
    (M_2)_{i_{12} i_{23} \dots i_{2K}}^{n_2}
    \dots (M_K)_{i_{1K} \dots i_{K-1\,K}}^{n_{K}}
    \ket{n_1 \dots n_{K}}\\
    % &= \sum_{
    %   \begin{smallmatrix}
    %     n_1 \dots n_K
    %   \end{smallmatrix}
    % } (M_1)^{n_1} \otimes \dots \otimes (M_K)^{n_{K}} \ket{n_1 \dots n_{K}}
  \end{split}
\end{equation}
where
$n_j$ is the occupation of the $j$\textsuperscript{th} spatial orbital,
$i_{jk}$ is an auxiliary index that represents the correlation of
$j$\textsuperscript{th} orbital with the $k$\textsuperscript{th} orbital,
and $M_j$ is a tensor that describes the correlation between the
$j$\textsuperscript{th} orbital and the other orbitals.
Each tensor is connected to others by at least one auxiliary index, meaning
that the correlation between orbitals is represented by tensor
contraction on the auxiliary indices.
The specific auxiliary indices used in the tensor-contraction control the
correlations that the wavefunction explicitly captures and thereby the
complexity of the wavefunction.
The Matrix Product State (MPS) wavefunction simplifies the TPS wavefunction by
only correlating the orbitals that are adjacent to each other in an ordered list
\cite{mps_dmrg,9,41,44,48}:
\begin{equation}
  \label{eq:mps}
  \begin{split}
    \ket{\Psi_{\mathrm{MPS}}} &= \sum_{
      \begin{smallmatrix}
        n_1, n_2, \dots, n_{K-1}, n_K\\
        i_{12}, i_{23}, \dots, i_{K-1\,K}
      \end{smallmatrix}
    } (M_1)_{i_{12}}^{n_1} (M_2)_{i_{12} i_{23}}^{n_2} \dots
    (M_K)_{i_{K-1\,K}}^{n_{K}} \ket{n_1 n_2 \dots n_{K-1} n_{K}}\\
    % &= \sum_{
    %   \begin{smallmatrix}
    %     n_1, n_2, \dots, n_{K-1}, n_K
    %   \end{smallmatrix}
    % }
    % (M_1)^{n_1} \otimes (M_2)^{n_2} \otimes \dots \otimes (M_{K-1})^{n_{K-1}} \otimes (M_K)^{n_{K}}
    % \ket{n_1 n_2 \dots n_{K-1} n_{K}}
  \end{split}
\end{equation}
MPS wavefunctions are usually optimized using the Density Matrix Renormalization
Group (DMRG) algorithm\cite{29, 30, 31, 32, 33, 34, 35, 36, 37, 38, dmrg_intro, ramonemail_dmrg}.

While MPS and TPS wavefunctions describe, in essence, the contribution of each
orbital to the wavefunction, the Antisymmetrized Product of Geminals (APG)
wavefunction describes the contribution of each electron pair (geminal) to the
wavefunction\cite{39, 40, 41, 42, 43, 44, 45, 46, 47, geminal_first, apg}:
\begin{equation}
  \label{eq:apg}
  \begin{split}
    \ket{\Psi_{\mathrm{APG}}}
    &= \prod_{p=1}^{N/2} G^\dagger_p \ket{0}\\
    G^\dagger_p \ket{0} &= \sum_{ij}^{2K} C_{p;ij} a_i^\dagger a_{j}^\dagger \ket{0}\\
  \end{split}
\end{equation}
where $G^\dagger_p$ is the creation operator of the $p$\textsuperscript{th}
geminal
and $C_{p;ij}$ is the contribution of the $i$\textsuperscript{th} and
$j$\textsuperscript{th} spin-orbitals to the $p$\textsuperscript{th} geminal.
Again, the complexity and the accuracy of this wavefunction can be controlled by
limiting the number of terms in the wavefunction.
For example, we can choose a set of orbital pairs that contribute to the
wavefunction.
The Antisymmetrized Product of Interacting Geminals (APIG) wavefunctions
and its variants\cite{48, 49, 50, 51, 52, 53, 54, 55, 56, 57, 58, 59, 60, 61,
  62, 63, 64, agp, 66, 67, 68, 69, 70, 71, 72, 73, 74, 75, apig_first}, such as
Antisymmetrized Product of 1-Reference Orbital Geminals (AP1roG)\cite{ap1rog} and
Antisymmetrized Product of Rank-2 Geminals (APr2G) wavefunctions\cite{apr2g}
only use spin-orbital pairs from the same spatial orbital:
\begin{equation}
  \label{eq:apig}
  \begin{split}
    \ket{\Psi_{\mathrm{APIG}}} &= \prod_p^{N/2} G_p^\dagger \ket{0}\\
    &= \prod_p^{N/2} \sum_{i}^{K} C_{p;i} a_i^\dagger a_{\bar{\imath}}^\dagger \ket{0}
  \end{split}
\end{equation}
where $a^\dagger_i$ and $a^\dagger_{\bar{\imath}}$ are the creation operators of the
alpha and beta spin-orbitals corresponding to the $i$\textsuperscript{th}
spatial orbital.

Each of these wavefunction ans\"{a}tze seems to be fundamentally different
in its nomenclature, structure, and computation.
Yet every wavefunction approximates the FCI wavefunction, and through this common
goal, they are intrinsically connected to one another.
There are known mathematical connections between certain ans\"{a}tze and these are
occasionally exploited to derive new flavours of these methods.
For example, many geminal methods can be rewritten as special CC
wavefunctions\cite{55, 56, 57, 59, ap1rog}.
However, these insights are seldom transferred between ans\"{a}tze and the
development of new ans\"{a}tze seems even rarer.
If the goal within electronic structure theory is to find the ansatz that
strikes the best balance between the cost and accuracy for a given system,
then do we not limit ourselves by committing to a particular ansatz and its
assumptions?
In this article, we present a \textit{general} wavefunction structure in which new
ans\"{a}tze can be easily developed and relations between existing ans\"{a}tze can be
elucidated.
We express popular existing multideterminant ans\"{a}tze (CI, CC, TPS, and APG
wavefunctions) within this framework and develop new structures by combining
their features.
We hope to demonstrate that wavefunction design can be reduced to unique
combinations of modular structures, indicating that an incredible number of ``new''
ans\"{a}tze can be trivially developed.

\section{Flexible Ansatz for N-particle Configuration Interaction (FANCI)}
\label{sec:FANCI}
The proposed wavefunction structure is quite simple and resembles the CI
wavefunction (Equation~\ref{eq:ci}):
\begin{equation}
  \label{eq:FANCI}
  \ket{\Psi_{\mathrm{FANCI}}} = \sum_{\mathbf{m} \in \setsd} f(\mathbf{m}, \vec{P}) \ket{\mathbf{m}}
\end{equation}
where $\setsd$ is a set of allowed Slater determinants and
$f$ is a function that controls the weight of each Slater determinant,
$\mathbf{m}$, using the parameters, $\vec{P}$.
Since Slater determinants can be uniquely represented with an excitation operator
and a reference, Equation~\ref{eq:FANCI} can be rewritten with respect to
excitation operators, $\hat{E}$.
\begin{equation}
  \label{eq:FANCI_excitation}
  \ket{\Psi_{\mathrm{FANCI}}} = \sum_{\hat{E} \in \setexc} f(\hat{E}, \vec{P}) \hat{E} \ket{\Phi_{\mathrm{ref}}}
\end{equation}
where $\setexc$ is a set of allowed excitation operators and $f$ is a
function that maps the weight of the excitation operator from $\hat{E}$ and
$\vec{P}$.
The $\setsd$ and $\setexc$ are equivalent representations of a set of Slater
determinants and can be used interchangeably.

In this paper, we aim to demonstrate the generality, utility, and flexibility of
this framework.
In the next section, we show that the framework is general by expressing popular ans\"{a}tze
as FANCI wavefunctions.
Then, in Section~\ref{sec:characteristics}, we discuss how the choices of $S$,
$\vec{P}$, and $f$ affect the accuracy, cost, and size-consistency of the wavefunction.
Finally, we demonstrate the flexibility of this framework by constructing novel
wavefunction structures.

\section{Examples}
\label{sec:examples}

\subsection{Hartree-Fock}
The ground-state Hartree-Fock (HF) wavefunction is the Slater determinant of
orthonormal orbitals that provides the lowest energy\cite{77, hf_?}.
Starting from an arbitrary set of orthonormal orbitals, created by
$\{a^\dagger_j\}$, the HF wavefunction can be obtained by optimizing the unitary
transformation that provides the lowest energy.
\begin{equation}
  \begin{split}
    \ket{\Psi_{\mathrm{HF}}}
    &= \prod_{i=1}^N \left( \sum_{j=1}^{2K} a^\dagger_j U_{ji} \right) \ket{0}\\
    &= \sum_{\mathbf{m}} |U(\mathbf{m})|^- \ket{\mathbf{m}}
  \end{split}
\end{equation}
where $U$ is a unitary matrix, and $U(\mathbf{m})$ is a submatrix of $U$
obtained by selecting the rows that correspond to the spin-orbitals in $\mathbf{m}$.
The derivation is given in the Appendix~\ref{sec:appendix_hf}.
If only $N$ orthonormal orbitals are rotated, or alternatively, if there is no
mixing of the occupied and virtual orbitals, then the HF wavefunction is
obtained trivially:
\begin{equation}
  \begin{split}
    \ket{\Psi_{\mathrm{HF}}}
    &= \prod_{i=1}^N \left( \sum_{j=1}^N a^\dagger_j U_{ji} \right) \ket{0}\\
    &= |U(\mathbf{m})|^- \ket{\mathbf{m}}\\
  \end{split}
\end{equation}
where $\mathbf{m}$ is the set of the occupied orbitals.
With normalization, $|U(\mathbf{m})|^-$ becomes $1$.
In other words, the HF wavefunction is invariant to rotation of the occupied
orbitals if there is no mixing between occupied and virtual orbitals
\cite{hf_invariance}.

\subsection{Truncated Configuration Interaction}
\label{sec:FANCI_ci}
The truncated CI wavefunction (Equation~\ref{eq:ci}) is a linear combination of
selected Slater determinants\cite{truncated_ci}.
Such wavefunctions can be trivially described in the proposed framework:
the set of allowed Slater determinants, $S$, is the same;
the parameters, $\vec{P}$, are the coefficients of the Slater determinants,
$\vec{C}$;
and the parameterizing function, $f$, simply selects the appropriate
coefficient, $C_\mathbf{m}$, given the Slater determinant, $\mathbf{m}$.
\begin{equation*}
  f(\mathbf{m}, \vec{C}) = \vec{e}_{\mathbf{m}} \cdot \vec{C}
\end{equation*}
where $\vec{e}_{\mathbf{m}}$ is a vector that gives $1$ in the
position of $\mathbf{m}$ and $0$ elsewhere.
Altogether, the CI wavefunction is
\begin{equation}
  \label{eq:FANCI_ci}
  \ket{\Psi_{\mathrm{CI}}} =
  \sum_{\mathbf{m} \in \setsd} \left( \vec{e}_{\mathbf{m}} \cdot \vec{C} \right) \ket{\mathbf{m}}
\end{equation}
If the CI wavefunction is expressed with respect to excitations on a reference,
we get
\begin{equation}
  \label{eq:FANCI_ci_exc}
  \ket{\Psi_{\mathrm{CI}}} =
  \sum_{\hat{E}_{\mathbf{i}}^{\mathbf{a}} \in \setexc}
  \left(
    \vec{e}_{\hat{E}_{\mathbf{i}}^{\mathbf{a}}} \cdot \vec{C}
  \right)
  \hat{E}_{\mathbf{i}}^{\mathbf{a}} \ket{\Phi_{ref}}
\end{equation}

\subsection{Coupled-Cluster}
\label{sec:FANCI_cc}
The CC wavefunction (Equation~\ref{eq:cc}) uses the exponential operator to
approximate high-order excitations as a product of lower-order excitations
\cite{ccsd}.
\begin{equation}
  \label{eq:cc2}
  \begin{split}
    \ket{\Psi_{\mathrm{CC}}} &= \exp(\hat{T}) \ket{\Phi_{\mathrm{HF}}}\\
    &= \sum_{n=0}^\infty \frac{1}{n!} \hat{T}^{n} \ket{\Phi_{\mathrm{HF}}}
  \end{split}
\end{equation}
where
\begin{equation}
  \label{eq:clusteroperator}
  \hat{T}
  = \sum_{
    \begin{smallmatrix}
      \hat{E}_{\mathbf{i}}^{\mathbf{a}} \in \tilde{S}_{\hat{E}}
    \end{smallmatrix}
  }
  t_{\mathbf{i}}^{\mathbf{a}} \hat{E}_{\mathbf{i}}^{\mathbf{a}}
\end{equation}
and $\tilde{S}_{\hat{E}}$ is a set of excitation operators.
The Maclaurin series in Equation~\ref{eq:cc2} lets one express the CI
coefficients in terms of CC cluster amplitudes $t_{\mathbf{i}}^{\mathbf{a}}$.
Specifically, the cluster amplitudes are cumulants of the CI coefficients
\cite{78, 79, 80, 81, 82}.
The powers of $\hat{T}$ (cluster operator) give the wavefunction access to
excitations beyond those allowed ($\tilde{S}_{\hat{E}}$) by generating all (product-wise)
combinations of the allowed excitation operators.
However, an excitation can be described with different combinations of
excitation operators, and the cumulant can be simplified by grouping together
terms that correspond to the same excitation (or Slater determinant).
Each combination corresponds to a subset of $\tilde{S}_{\hat{E}}$, such that the set of all Slater
determinants in the CC wavefunction can be described in terms of all possible
subsets of $\tilde{S}_{\hat{E}}$:
\begin{equation}
  \label{eq:FANCI_cc_set}
  \setexc = \left\{ \prod_{\hat{E}_k \in T} \hat{E}_k \middle| T \subseteq \tilde{S}_{\hat{E}} \right\}
\end{equation}
Then, the wavefunction can be written as a sum over all possible Slater
determinants and a sum over all possible combinations of excitation operators
that produce the given Slater determinant.
\begin{equation}
  \label{eq:FANCI_cc_f}
  \begin{split}
    f(\hat{E}_{\mathbf{i}}^{\mathbf{a}}, \mathbf{t}) &=
    \mathlarger{\sum}_{
      \begin{smallmatrix}
        \{\hat{E}_{\mathbf{i}_1}^{\mathbf{a}_1} \dots \hat{E}_{\mathbf{i}_n}^{\mathbf{a}_n}\}
        \subseteq \tilde{S}_{\hat{E}}\\
        \mathrm{sgn} \prod_{k=1}^n \hat{E}_{\mathbf{i}_k}^{\mathbf{a}_k} = \hat{E}_{\mathbf{i}}^{\mathbf{a}}
      \end{smallmatrix}
    }
    \mathrm{sgn}(\sigma_{\hat{E}_{\mathbf{i}_1}^{\mathbf{a}_1} \dots \hat{E}_{\mathbf{i}_n}^{\mathbf{a}_n}})
    \frac{1}{n!}
    \begin{vmatrix}
      t_{\mathbf{i}_1}^{\mathbf{a}_1} & \dots & t_{\mathbf{i}_n}^{\mathbf{a}_n}\\
      \vdots & \ddots & \vdots\\
      t_{\mathbf{i}_1}^{\mathbf{a}_1} & \dots & t_{\mathbf{i}_n}^{\mathbf{a}_n}
    \end{vmatrix}^+\\
    &=
    \mathlarger{\sum}_{
      \begin{smallmatrix}
        \{\hat{E}_{\mathbf{i}_1}^{\mathbf{a}_1} \dots \hat{E}_{\mathbf{i}_n}^{\mathbf{a}_n}\}
        \subseteq \tilde{S}_{\hat{E}}\\
        \mathrm{sgn} \prod_{k=1}^n \hat{E}_{\mathbf{i}_k}^{\mathbf{a}_k} = \hat{E}_{\mathbf{i}}^{\mathbf{a}}
      \end{smallmatrix}
    }
    \mathrm{sgn}(\sigma_{\hat{E}_{\mathbf{i}_1}^{\mathbf{a}_1} \dots \hat{E}_{\mathbf{i}_n}^{\mathbf{a}_n}})
    \mathlarger{\prod}_{k=1}^n t_{\mathbf{i}_k}^{\mathbf{a}_k}
  \end{split}
\end{equation}
where $n$ is the dimension of the subset
$\{\hat{E}_{\mathbf{i}_1}^{\mathbf{a}_1} \dots \hat{E}_{\mathbf{i}_n}^{\mathbf{a}_n}\}$.
The sum can be interpreted as a sum over all possible partitions of a given
excitation operator, $\hat{E}_{\mathbf{i}}^{\mathbf{a}}$, into excitations from
the given set.
The signature of the permutation,
$\mathrm{sgn}(\sigma_{\hat{E}_{\mathbf{i}_1}^{\mathbf{a}_1} \dots \hat{E}_{\mathbf{i}_n}^{\mathbf{a}_n}})$,
results from reordering the creation and annihilation operators of the
lower-order excitations to the same order as the given excitation operator:
\begin{equation}
  \hat{E}_{\mathbf{i}}^{\mathbf{a}} =
  \mathrm{sgn}(\sigma_{\hat{E}_{\mathbf{i}_1}^{\mathbf{a}_1} \dots \hat{E}_{\mathbf{i}_n}^{\mathbf{a}_n}})
  \hat{E}_{\mathbf{i}_1}^{\mathbf{a}_1} \cdots \hat{E}_{\mathbf{i}_n}^{\mathbf{a}_n}
\end{equation}
The permanent, $|A|^+$, accounts for all possible orderings within a given set of
excitation operators.
Altogether, the CC wavefunction can be reformulated as
\begin{equation}
  \label{eq:FANCI_cc}
  \ket{\Psi_{\mathrm{CC}}} =
  \mathlarger{\sum}_{
    \hat{E}_{\mathbf{i}}^{\mathbf{a}} \in \setexc
  }
  \;
  \left(
    \mathlarger{\sum}_{
      \begin{smallmatrix}
        \{\hat{E}_{\mathbf{i}_1}^{\mathbf{a}_1} \dots \hat{E}_{\mathbf{i}_n}^{\mathbf{a}_n}\}
        \subseteq \tilde{S}_{\hat{E}}\\
        \mathrm{sgn} \prod_{k=1}^n \hat{E}_{\mathbf{i}_k}^{\mathbf{a}_k} = \hat{E}_{\mathbf{i}}^{\mathbf{a}}
      \end{smallmatrix}
    }
    \mathrm{sgn}(\sigma_{\hat{E}_{\mathbf{i}_1}^{\mathbf{a}_1} \dots \hat{E}_{\mathbf{i}_n}^{\mathbf{a}_n}})
    \mathlarger{\prod}_{k=1}^n t_{\mathbf{i}_k}^{\mathbf{a}_k}
  \right)
  \hat{E}_{\mathbf{i}}^{\mathbf{a}} \ket{\Phi_{\mathrm{HF}}}
\end{equation}
More details are provided in the Appendix~\ref{sec:appendix_cc}.

\subsection{Tensor Product State}
\label{sec:FANCI_tps}
The TPS (and MPS) wavefunction (Equation~\ref{eq:tps}) determines the weight of
a Slater determinant by tensor (and matrix) contractions, where
each shared index corresponds to a correlation between
orbitals\cite{tns_intro, dmrg_intro}:
\begin{equation*}
  \begin{split}
    \ket{\Psi_{\mathrm{TPS}}} &= \sum_{
      \begin{smallmatrix}
        n_1 \dots n_K
      \end{smallmatrix}
    } \sum_{
      \begin{smallmatrix}
        i_{12} \dots i_{1K}\\
        i_{23} \dots i_{2K}\\
        \vdots\\
        i_{K-1\, K}
      \end{smallmatrix}
    } (M_1)_{i_{12} \dots i_{1K}}^{n_1}
    (M_2)_{i_{12} i_{23} \dots i_{2K}}^{n_2}
    \dots (M_K)_{i_{1K} \dots i_{K-1\,K}}^{n_{K}}
    \ket{n_1 \dots n_{K}}\\
    % &= \sum_{
    %   \begin{smallmatrix}
    %     n_1 \dots n_K
    %   \end{smallmatrix}
    % } (M_1)^{n_1} \otimes \dots \otimes (M_K)^{n_{K}} \ket{n_1 \dots n_{K}}
  \end{split}
\end{equation*}
Each spatial orbital, $k$, is associated with a tensor, $M_k$, and each tensor
is associated with the occupation of its spatial orbital (i.e. its
state), $n_k$, and with other tensors using its auxiliary indices,
$\{i_{1k} \dots i_{k-1\,k}\; i_{k\, k+1} \dots i_{kK}\}$.
Then, the coefficient associated with the Slater determinant, represented by
$\{n_1 \dots n_K\}$, is approximated by tensor-contraction.
Many variants of TPS, including MPS, impose some structure on the
tensor product so that the evaluation and optimization of the wavefunction are
computationally tractable.

Since $\{n_1 \dots n_K\}$ is yet another representation of the Slater
determinant, we can describe the wavefunction with respect to $\mathbf{n}$:
\begin{equation*}
  \mathbf{n} = \{n_1 \dots n_K\}
\end{equation*}
Therefore, the TPS wavefunction can be rewritten as
\begin{equation}
  \label{eq:FANCI_tps}
  \begin{split}
    \ket{\Psi_{\mathrm{TPS}}} &= \sum_{
      \mathbf{n} \in S_{\mathrm{FCI}}
    }
    \bigodot_{k=1}^K (M_k)^{n_k}
    \ket{\mathbf{n}}\\
  \end{split}
\end{equation}
where $K$ is the number of spatial orbitals, and
$\bigodot$ describes the specific tensor-contraction used in the wavefunction.
While it is not common to do so, the TPS wavefunctions can be
equivalently expressed with respect to spin-orbitals.
If each state of the TPS wavefunction corresponds to the occupation of a
spin-orbital, $m_k$, then the same notation can be used as in
Equation~\ref{eq:FANCI}
\begin{equation}
  \label{eq:FANCI_tps_m}
  \begin{split}
    \ket{\Psi_{\mathrm{TPS}}} &= \sum_{
      \mathbf{m} \in S_{\mathrm{FCI}}
    }
    \bigodot_{k=1}^{2K} (M_k)^{m_k}
    \ket{\mathbf{m}}\\
  \end{split}
\end{equation}
where $M_k$ is the tensor associated with spin-orbital $k$.

\subsection{Antisymmetrized Product of Geminals}
\label{sec:FANCI_apg}
Just as the HF wavefunction is constructed as an antisymmetrized product of
one-electron wavefunctions (orbitals), the APG wavefunction is constructed as an
antisymmetrized product of two-electron wavefunctions (geminals)\cite{39, 42,
  43, apg}:
% (FIXME)
\begin{equation*}
  \begin{split}
    \ket{\Psi_{\mathrm{APG}}}
    &= \prod_{p=1}^{N/2} G^\dagger_p \ket{0}\\
    &= \prod_{p=1}^{N/2} \sum_{ij}^{2K} C_{p;ij} a_i^\dagger a_{j}^\dagger \ket{0}\\
    &= \prod_{p=1}^{N/2} \sum_{\mathbf{m}_k} C_{p;\mathbf{m}_k} A^\dagger_{\mathbf{m}_k} \ket{0}\\
  \end{split}
\end{equation*}
where the $\mathbf{m}_k$ denotes a set of a pair of indices and
$A^\dagger_{\mathbf{m}_k}$ denotes the creation operator that corresponds to
$\mathbf{m}_k$.
Similar to the CC wavefunction, the product of sums can be expanded out as a
sum over each Slater determinant and a sum over the different combinations of
electron pairs that create the Slater determinant.
In the HF wavefunction, the product of sums results in a determinant due to the
antisymmetry with respect to the interchange of electrons.
In the APG wavefunction, however, the interchange of electron pairs is symmetric,
and the product of sums results in a permanent.
Given the set of all possible two-electron creation operators, $\tilde{S}$, a subset of
exactly $\frac{N}{2}$ two-electron creators,
$\{A^\dagger_{\mathbf{m}_1} \dots A^\dagger_{\mathbf{m}_{N/2}}\}$,
is needed to construct a given Slater determinant, $\mathbf{m}$, where the
number of electrons, $N$, is even.
Since an orbital cannot be occupied more than once and all the orbitals are
necessary to construct a given Slater determinant, any selection of orbital
pairs, $\{\mathbf{m}_1 \dots \mathbf{m}_{N/2}\}$, must be disjoint and exhaustive.
\begin{equation}
  \label{eq:FANCI_apg_f}
  \begin{split}
    f(\mathbf{m}, \mathbf{C}) &=
    \mathlarger{\sum}_{
      \begin{smallmatrix}
        \{\mathbf{m}_1 \dots \mathbf{m}_{N/2}\} \subseteq \tilde{S}\\
        \bigcup\limits_{k=1}^{N/2} \mathbf{m}_k = \mathbf{m}\\
        \mathbf{m}_p \cap \mathbf{m}_q = \emptyset \; \forall p \neq q
      \end{smallmatrix}
    }
    \mathrm{sgn}(\sigma_{\mathbf{m}_1 \dots \mathbf{m}_{N/2}}) |C(\mathbf{m}_1, \dots, \mathbf{m}_{N/2})|^+\\
    &=
    \mathlarger{\sum}_{
      \begin{smallmatrix}
        \{\mathbf{m}_1 \dots \mathbf{m}_{N/2}\} \subseteq \tilde{S}\\
        \mathrm{sgn} A^\dagger_{\mathbf{m}_1} \dots A^\dagger_{\mathbf{m}_{N/2}} \ket{0} = \ket{\mathbf{m}}
      \end{smallmatrix}
    }
    \mathrm{sgn}(\sigma_{\mathbf{m}_1 \dots \mathbf{m}_{N/2}}) |C(\mathbf{m}_1, \dots, \mathbf{m}_{N/2})|^+
  \end{split}
\end{equation}
where the orbital pairs, $\{\mathbf{m}_1 \dots \mathbf{m}_{N/2}\}$, are selected
such that they result in the given Slater determinant, $\mathbf{m}$, without
duplicate orbitals.
Similar to the CC wavefunction (Equation~\ref{eq:FANCI_cc_f}), the sum
can be interpreted as a sum over all allowed partitions of the given Slater
determinant into the electron pairs.
The signature of the permutation, $\mathrm{sgn}(\sigma_{\mathbf{m}_1 \dots \mathbf{m}_{N/2}})$,
results from reordering the creation operators in the electron pairs to the same
order as in the given Slater determinant:
\begin{equation}
  \prod_{i \in \mathbf{m}} a^\dagger_i =
  \mathrm{sgn}(\sigma_{\mathbf{m}_1 \dots \mathbf{m}_{N/2}})
  \prod_{p=1}^{N/2} A^\dagger_{\mathbf{m}_p}
\end{equation}
Altogether, the APG wavefunction is reformulated as
\begin{equation}
  \label{eq:FANCI_apg}
  \negtab
  \ket{\Psi_{\mathrm{APG}}} =
  \sum_{\mathbf{m} \in \setsd^{\mathrm{FCI}}}
  \left(
    \mathlarger{\sum}_{
      \begin{smallmatrix}
        \{\mathbf{m}_1 \dots \mathbf{m}_{N/2}\} \subseteq \tilde{S}\\
        \mathrm{sgn} A^\dagger_{\mathbf{m}_1} \dots A^\dagger_{\mathbf{m}_{N/2}} \ket{0} = \ket{\mathbf{m}}
      \end{smallmatrix}
    }
    \mathrm{sgn}(\sigma_{\mathbf{m}_1 \dots \mathbf{m}_{N/2}}) |C(\mathbf{m}_1, \dots, \mathbf{m}_{N/2})|^+
  \right)
  \ket{\mathbf{m}}
\end{equation}
where $\setsd^{\mathrm{FCI}}$ is a set of all possible Slater determinants (i.e.
Slater determinants of a FCI wavefunction).
The derivation is given in the Appendix~\ref{sec:appendix_apg}.

The Antisymmetrized Product of Interacting Geminals (APIG) is a special case of
the APG wavefunction such that only the electron pairs within the same spatial
orbital, i.e. doubly occupied spatial orbitals, are used to build the
wavefunction\cite{apig_first}.
The sum over the partitions reduces to a single element because there is only
one way to construct a given (seniority-zero) Slater determinant from electron
pairs of doubly occupied orbitals.
\begin{equation}
  \label{eq:FANCI_apig}
  \ket{\Psi_{\mathrm{APIG}}} =
  \sum_{\mathbf{m} \in \setsd^{\mathrm{DOCI}}} |C(\mathbf{m})|^+ \ket{\mathbf{m}}
\end{equation}
where $\setsd^{\mathrm{DOCI}}$ is the set of all seniority-zero (no unpaired
electrons) Slater determinants, and
$|C(\mathbf{m})|^+$ is a permanent of the parameters that correspond to the
spatial orbitals used to construct $\mathbf{m}$.
The APIG wavefunction can be further simplified by imposing structures onto the
permanent:
the Antisymmetrized Product of 1-reference Orbitals Geminals (AP1roG) wavefunction
assumes that a large portion of the coefficient matrix is an identity matrix
\cite{ap1rog};
and the Antisymmetrized Product of rank-2 Geminals (APr2G) wavefunction assumes
that the coefficient matrix is a Cauchy matrix\cite{apr2g}.
APr2G reduces the cost of evaluating a permanent ($\mathcal{O}(n!)$) to that of
a determinant ($\mathcal{O}(n^3)$).
AP1roG has the cost of $\mathcal{O}(m!)$ where $m$ is the order of excitation
with respect to the reference Slater determinant.
It is cheap to evaluate the overlap of the AP1roG wavefunction with low-order
excitations of the reference determinant.

\subsection{Universality of FANCI}
The CI, CC, TPS, and APG wavefunctions and their variants can be
expressed within the FANCI framework using different $S$, $\vec{P}$, and $f$.
We can define a multideterminant wavefunction as a function that has a
well-defined overlap with a set of orthonormal Slater determinants.
Provided that the wavefunction exists within the space spanned by the Slater
determinants, %(FIXME)
the wavefunction can be re-expressed as a linear
combination of Slater determinants via a projection:
\begin{equation}
  \begin{split}
    \ket{\Psi (\vec{P})} &=
    \sum_{\mathbf{m} \in \setsd} \ket{\mathbf{m}} \braket{\mathbf{m} | \Psi (\vec{P})}\\
    &= \sum_{\mathbf{m} \in \setsd} f(\mathbf{m}, \vec{P}) \ket{\mathbf{m}}
  \end{split}
\end{equation}
where
\begin{equation}
  f(\mathbf{m}, \vec{P}) = \Braket{\mathbf{m} | \Psi (\vec{P})}
\end{equation}
Therefore, all multideterminant wavefunctions, as defined above, can be
expressed within the framework of Equation~\ref{eq:FANCI}:
$S$ is the minimal set of Slater determinants required to fully describe the
wavefunction;
$\vec{P}$ is the parameters of the wavefunction;
and $f$ is the overlap of the wavefunction with the Slater determinant,
$\mathbf{m}$.

\section{Characteristics}
\label{sec:characteristics}
In the formulation of Equation~\ref{eq:FANCI}, a multideterminant wavefunction
is defined using only a specified (sub)set of Slater determinants, $S$,
wavefunction parameters, $\vec{P}$, and function $f$.
Since the characteristics of a wavefunction ansatz depend on its structure,
all characteristics of a multideterminant wavefunction can be deduced from the
specified $S$, $\vec{P}$, and $f$.
Designing a wavefunction with desirable characteristics, therefore, merely
requires selecting $S$, $\vec{P}$, and $f$.
We propose to approach method development in electronic structure theory as a
search for $S$, $\vec{P}$, and $f$ that induce the desired wavefunction
features.
While many features can be considered, and we shall consider additional features in
future work, here we shall address just three important characteristics:
accuracy, cost, and size-consistency.

\subsection{Accuracy}
\label{sec:accuracy}
Ultimately, the FANCI wavefunction models the FCI wavefunction by
parameterizing the weights of each Slater determinant.
If there are Slater determinants absent from the FANCI wavefunction, i.e. $S
\subset S_{\mathrm{FCI}}$, then the omitted Slater determinants are assumed to
have no contributions to the FCI wavefunction.
The effects of $S$ can be viewed as a modification of the parameterizing
function.
\begin{equation*}
  \ket{\Psi_{\mathrm{FANCI}}} =
  \sum_{\mathbf{m} \in S_{\mathrm{FCI}}} g(\mathbf{m}, \vec{P}) \ket{\mathbf{m}}
\end{equation*}
where
\begin{equation*}
  g(\mathbf{m}, \vec{P}) =
  \begin{cases}
    f(\mathbf{m}, \vec{P}) & ; \mathbf{m} \in S\\
    0 & ; \mathbf{m} \not\in S\\
  \end{cases}
\end{equation*}
Alternatively, the FANCI wavefunction can be viewed as a model for the CI
wavefunction built using the same restricted set of Slater determinants.
In either case, preventing Slater determinants from contributing to the
wavefunction will cause deviations from the FCI wavefunction.

As with any parameterization (or fitting) problem, it becomes easier to
find a function that accurately describes each weight as the number of
parameters increase.
FANCI wavefunctions cannot be exact, in general, unless the number of parameters
is greater than or equal to the number of parameters in the Hamiltonian which,
in the case of electronic structure theory, means there should be at least as
many parameters as there are two-electron integrals\cite{83, 84}.
Methods with many fewer parameters are, typically, static correlation methods.
On the other hand, appropriately constructed FANCI ans\"{a}tze should approach the
FCI limit as the number of parameters approaches the number of Slater
determinants.
However, the cost associated with optimizing the wavefunction typically
increases superlinearly as the number of parameters increases.

\subsection{Cost}
\label{sec:cost}
The cost associated with a wavefunction can be divided into the cost of its
storage, evaluation, and optimization, all of which are intricately linked.
The cost of storage is associated with the number of parameters needed to
describe the wavefunction.
The cost of evaluating the wavefunction depends on the cost of evaluating
$f$ and on the number of times $f$ needs to be evaluated.
% TODO: move to separate section (optimization?)
For example, in order to evaluate the norm of a wavefunction, $f$ must be
evaluated for every Slater determinant in $S$.
\begin{equation*}
  \begin{split}
    \braket{\Psi | \Psi}
    &= \sum_{\mathbf{m} \in S} \sum_{\mathbf{n} \in S}
    f^*(\mathbf{m}) \braket{\mathbf{m} | \mathbf{n}} f(\mathbf{n})\\
    &= \sum_{\mathbf{m} \in S} f^*(\mathbf{m}) f(\mathbf{m})
  \end{split}
\end{equation*}
Upon optimization, a new set of parameters are found such that the wavefunction
satisfies the Schr\"{o}dinger equation.
\begin{equation}
  \label{eq:classic_schr}
  \hat{H} \ket{\Psi} = E \ket{\Psi}
\end{equation}
Equation~\ref{eq:classic_schr} is often rewritten in its variational form or
its projected form to make it easier to solve numerically.
The optimization procedure and the associated costs depend on the equations that
are being solved.

The variational Schr\"{o}dinger equation involves integrating both sides of
Equation~\ref{eq:classic_schr} with the wavefunction\cite{fci}.
\begin{equation}
  \label{eq:var_schr_FANCI}
  \begin{split}
    \braket{\Psi  | \hat{H} | \Psi} &= E \braket{\Psi | \Psi}\\
    \sum_{\mathbf{m}, \mathbf{n} \in S} f^*(\mathbf{m},\vec{P})
    \braket{\mathbf{m} | \hat{H} | \mathbf{n}} f(\mathbf{n},\vec{P}) &= E
    \sum_{\mathbf{m} \in S} f^*(\mathbf{m},\vec{P}) f(\mathbf{m},\vec{P})\\
  \end{split}
\end{equation}
If the number of Slater determinants in $S$ is comparable to those in the FCI
wavefunction, even setting up Equation~\ref{eq:var_schr_FANCI} will
require far too many evaluations of $f$ to be computationally tractable.
During the optimization, all terms need to be evaluated at each step, where the
number of steps needed for convergence varies depending on the system and the
optimization algorithm.

The projected Schr\"{o}dinger equation can be obtained from Equation~\ref{eq:classic_schr} using the resolution of identity:
\begin{equation}
  \begin{split}
    \hat{H} \ket{\Psi} &= E \ket{\Psi}\\
    \left(
      \sum_{\mathbf{m} \in S_{\mathrm{FCI}}} \ket{\mathbf{m}} \bra{\mathbf{m}}
    \right)
    \hat{H} \ket{\Psi} &= E
    \left(
      \sum_{\mathbf{m} \in S_{\mathrm{FCI}}} \ket{\mathbf{m}} \bra{\mathbf{m}}
    \right)
    \ket{\Psi}\\
    \sum_{\mathbf{m} \in S_{\mathrm{FCI}}} 
    \ket{\mathbf{m}}
    \left(
        \braket{\mathbf{m} | \hat{H} | \Psi}
        - E \braket{\mathbf{m} | \Psi}
    \right)
    &= 0\\
  \end{split}
\end{equation}
Since the Slater determinants are linearly independent, this equation will hold only if $\braket{\mathbf{m} | \hat{H} | \Psi} = E\braket{\mathbf{m} | \Psi}$ for every Slater determinant $\mathbf{m}$.
These equations are expressed as a system of equations:
\begin{equation}
  \label{eq:proj_schr_2}
  \begin{split}
    \braket{\mathbf{m}_1 | \hat{H} | \Psi} - E \braket{\mathbf{m}_1 | \Psi} &= 0\\
    &\hspace{0.5em}\vdots\\
    \braket{\mathbf{m}_M | \hat{H} | \Psi} - E\braket{\mathbf{m}_M | \Psi} &= 0\\
  \end{split}
\end{equation}

Essentially, the Schr\"{o}dinger equation (Equation~\ref{eq:classic_schr}) is
broken apart into separate equations for each contributing Slater determinant.
If the projection operator is not complete (i.e. contributions from certain
Slater determinants are discarded) then the equation (or system of equations)
will be an approximation of the original Equation~\ref{eq:classic_schr}.
% Since the Hamiltonian only contains the first and second order excitation
% operators, the weights of some Slater determinants ($S$) will not be evaluated
% throughout the optimization if enough equations are discarded.
% However, with an appropriate parameterization, these weights will not
% necessarily be zero because the weight

In general, the Schr\"{o}dinger equation can be expressed with respect to arbitrary function, $\Phi$\cite{59}.
\begin{equation}
  \label{eq:arb_schr}
  \braket{\Phi | \hat{H} | \Psi} = E \braket{\Phi | \Psi}\\
\end{equation}
If $\Phi$ is not $\Psi$, then certain components of $\Psi$ may be
projected out, imposing additional structure on the wavefunction through the
optimization process.
We can express this projection explicitly with a projection operator onto a
set of basis functions.
When this basis set is complete, the projected Schr\"{o}dinger equation is equivalent to the variational formulation.

Therefore, we can interpret the projected Schr\"{o}dinger equation as an approximation to the variational formulation that reduces it to a set of functions that capture the important
characteristics of the wavefunction.
The cost of evaluating Equation~\ref{eq:proj_schr_2} and \ref{eq:arb_schr}
depends on the functions onto which the Schr\"{o}dinger equation is projected.
Some wavefunction structures have special functions such that
Equation~\ref{eq:proj_schr_2} or \ref{eq:arb_schr} can be evaluated cheaply.
For example, the CC wavefunctions are often projected against
$\bra{\Phi_{\mathrm{HF}}} \exp(-\hat{T})$ because
$\braket{\Phi_{\mathrm{HF}} | \exp(-\hat{T}) \hat{H} \exp(\hat{T}) | \Phi_{\mathrm{HF}}}$,
can be simplified using the Baker-Campbell-Hausdorff expansion\cite{size-consistency}.
For a general FANCI wavefunction, however, it is convenient to project onto
a set of Slater determinants, $\{ \mathbf{m}_1 \dots \mathbf{m}_M\}$,
obtaining a system of (generally nonlinear) equations to solve:
\begin{equation}
  \label{eq:proj_schr_FANCI}
  \begin{split}
    \sum_{\mathbf{m} \in S}
    f(\mathbf{m},\vec{P}) \braket{\mathbf{m}_{1} | \hat{H} | \mathbf{m}} &= E
    f(\mathbf{m}_{1},\vec{P})\\
    &\hspace{0.5em}\vdots\\
    \sum_{\mathbf{m} \in S}
    f(\mathbf{m},\vec{P}) \braket{\mathbf{m}_{M} | \hat{H} | \mathbf{m}} &= E
    f(\mathbf{m}_{M},\vec{P})\\
  \end{split}
\end{equation}
In order to find a solution, the number of equations in the system of equations
must be greater than the number of unknowns.
Since the number of possible Slater determinants grows exponentially, there will
not be a shortage of equations (Slater determinants), and the number of
equations will almost always be greater than the number of unknowns.
If the least-squares solution of the nonlinear equations is found, then the
residual can be used to measure the error associated with the optimized
wavefunction (and energy).

Unless there is a special algorithm that limits the number of evaluated terms in
the variational Schr\"{o}dinger equation (Equation~\ref{eq:var_schr_FANCI}) or
a function $\Phi$ that allows cheap integration of the Schr\"{o}dinger equation
(Equation~\ref{eq:arb_schr}), a wavefunction should be evaluated using the
projected Schr\"{o}dinger equation (Equation~\ref{eq:proj_schr_2}) to control the optimization process.
Both the cost and accuracy of the wavefunction can be controlled;
as the number of projections increases, both accuracy and cost increases.
In addition, we can impose symmetry on the wavefunction by projecting the
Schr\"{o}dinger equation onto a space that satisfies a particular symmetry
\cite{73, proj_var, ramonemail_sym1, ramonemail_sym2, ramonemail_sym3}.
For example, we can reintroduce particle number symmetry onto a number-symmetry
broken wavefunction by projecting it onto Slater determinants with the selected
particle number.

\subsection{Size-Consistency}
\label{sec:sizeconsistency}
Let there be a system, $AB$, composed of two non-interacting subsystems, $A$ and
$B$.
Then, a wavefunction is size-consistent if the energy of the wavefunction
for $AB$ is the sum of the energies of the wavefunctions for $A$ and $B$,
\cite{size-consistency} i.e.
\begin{equation}
  \label{eq:sizeconsistency_dfn}
  \begin{split}
    H_{AB} \ket{\Psi_{AB}}
    &= E_{AB} \ket{\Psi_{AB}}\\
    &= (E_A + E_B) \ket{\Psi_{AB}}\\
  \end{split}
\end{equation}
where
\begin{equation}
  \begin{split}
    H_A \ket{\Psi_A} &= E_A \ket{\Psi_A}\\
    H_B \ket{\Psi_B} &= E_B \ket{\Psi_B}
  \end{split}
\end{equation}
Since subsystems $A$ and $B$ are non-interacting, there are no nonzero terms in
the Hamiltonian that couple $A$ and $B$, i.e. $H_{AB} = H_A + H_B$.
Then, the (partially symmetry broken) wavefunction $\Psi_{AB}$ can be written as
a product of the (orthogonal) subsystem wavefunctions, $\Psi_A$ and $\Psi_B$.
\begin{equation*}
  \ket{\Psi_{AB}} = \ket{\Psi_A} \ket{\Psi_B}
\end{equation*}
% TODO: add Ramon's size consistency proof

Similarly, a FANCI wavefunction is size-consistent if the weight function is
multiplicatively separable:
\begin{equation}
  \label{eq:FANCI_sep}
  \begin{split}
    \ket{\Psi_{AB}}
    &= \sum_{\mathbf{m} \in S_{AB}} f_{AB}(\mathbf{m}, \vec{P}) \ket{\mathbf{m}}\\
    % &= \sum_{\mathbf{m} \in S_{AB}} f_{A}(\mathbf{m}, \vec{P}) f_{B}(\mathbf{m}, \vec{P}) \ket{\mathbf{m}}\\
    &= \sum_{\mathbf{m}_A \in S_A} \sum_{\mathbf{m}_B \in S_B} f_{A}(\mathbf{m}_A, \vec{P}_A)
    f_{B}(\mathbf{m}_B, \vec{P}_B) \ket{\mathbf{m}_A} \ket{\mathbf{m}_B}\\
    &= \sum_{\mathbf{m}_A \in S_A} f_{A}(\mathbf{m}_A, \vec{P}_A) \ket{\mathbf{m}_A}
    \sum_{\mathbf{m}_B \in S_B} f_{B}(\mathbf{m}_B, \vec{P}_B) \ket{\mathbf{m}_B}\\
    &= \ket{\Psi_A} \ket{\Psi_B}
  \end{split}
\end{equation}
where subscripts $A$ and $B$ designate that the quantity belongs only to
subsystem $A$ and $B$, respectively.
This not only requires that $f$ be multiplicatively separable $f$, i.e.
$f_{AB} = f_A f_B$,
but also that the Slater determinants, $\mathbf{m}$, and the parameters,
$\vec{P}$, must be divisible into two disjoint parts,
$\{\mathbf{m}_A, \mathbf{m}_B\}$ and $\{\vec{P}_A, \vec{P}_B\}$ respectively.
To separate each Slater determinant into the subsystems, $\mathbf{m}$ must be
expressed using orbitals localized to each subsystem.
Since each $\mathbf{m}$ can have varying contribution from the subsystems $A$
and $B$, $S_A$ and $S_B$ contain Slater determinants with different numbers of
electrons.
However, we can impose the particle number symmetry during the optimization process.
Similarly, the parameters must represent quantities that are specific to each
subsystem.
Notice that $f_{AB} = f_A f_B$ is true when $f$ is a determinant (Hartree-Fock),
exponential (Coupled-Cluster), or permanent (geminals).
Note that this is a simplified approach to size-consistency and a more rigorous approach requires ensuring that the systems have the correct symmetries upon dissociation.

\section{Ans\"{a}tze}
\label{sec:new_ansatze}
% TODO: Add size-consistency and FCI limit
In the formulation of Equation~\ref{eq:FANCI}, a multideterminant wavefunction
can be entirely expressed with a set of Slater determinants, $S$, parameters,
$\vec{P}$, and a weight function, $f$.
Altering the $S$, $\vec{P}$, or $f$ of a given ansatz will effectively result in a
new ansatz.
Additionally, the optimization method can be modified to produce an ``ansatz''
with a different accuracy and cost.
For example, DMRG is an algorithm for optimizing MPS\cite{85}.
Here, we modify the FANCI forms of the CC (Equation~\ref{eq:FANCI_cc}), TPS
(Equation~\ref{eq:FANCI_tps}), and APG (Equation~\ref{eq:FANCI_apg})
wavefunctions to construct several new wavefunction structures.

\subsection{CC with Creators}
\label{sec:cc_variants}
Just as the TPS and APG wavefunctions are expressed with respect to creation
operators, we can replace the excitation operators in the CC wavefunction with
creation operators.
\begin{equation}
  \ket{\Psi_{\mathrm{CC}}} = \exp
  \left(
    \sum_{\mathbf{b}_i} C_{\mathbf{b}_i} \hat{A}^\dagger_{\mathbf{b}_i} +
    \sum_{\mathbf{f}_i} C_{\mathbf{f}_i} \hat{A}^\dagger_{\mathbf{f}_i}
  \right)
  \ket{0}
\end{equation}
where $A_{\mathbf{b}_i}^\dagger$ is a creation operator of even number of
electrons (denoted as even-electron),
$A_{\mathbf{f}_i}^\dagger$ is a creation operator of odd number of electrons
(denoted as odd-electron),
$\mathbf{b}_i$ is the set of orbitals created by $A^\dagger_{\mathbf{b}_i}$, and
$\mathbf{f}_i$ is the set of orbitals created by $A^\dagger_{\mathbf{f}_i}$.
For a consistent notation, we define $\tilde{S}_{\mathbf{b}}$ and
$\tilde{S}_{\mathbf{f}}$ as the set of allowed creation operators.
Since each creation operator can create a different number of electrons, the
total number of operators needed for $\mathbf{m}$ can vary depending on the
selection of creation operators.
For a given Slater determinant, let $n_b$ be the number of even-electron creators
and $n_f$ be the number of odd-electron creators.
Similar to the APG wavefunction, there can be multiple combinations of creation
operators that give the same Slater determinant.
We represent these combinations as a sum over
$\{ \mathbf{b}_1 \dots \mathbf{b}_{n_b} \mathbf{f}_1 \dots \mathbf{f}_{n_f} \}$
such that the product of the associated creators results in the given Slater
determinant:
\begin{equation}
  \prod_{i \in \mathbf{m}} a^\dagger_i
  =
  \mathrm{sgn}
  \left(
    \sigma(
    \hat{A}^\dagger_{\mathbf{b}_1} \dots \hat{A}^\dagger_{\mathbf{b}_{n_b}}
    \hat{A}^\dagger_{\mathbf{f}_1} \dots \hat{A}^\dagger_{\mathbf{f}_{n_f}}
    )
  \right)
  \hat{A}^\dagger_{\mathbf{b}_1} \dots \hat{A}^\dagger_{\mathbf{b}_{n_b}}
  \hat{A}^\dagger_{\mathbf{f}_1} \dots \hat{A}^\dagger_{\mathbf{f}_{n_f}}
\end{equation}
Similar to the signature in the CC wavefunction,
$ \mathrm{sgn}
\left(
  \sigma(
  \hat{A}^\dagger_{\mathbf{b}_1} \dots \hat{A}^\dagger_{\mathbf{b}_{n_b}}
  \hat{A}^\dagger_{\mathbf{f}_1} \dots \hat{A}^\dagger_{\mathbf{f}_{n_f}}
  )
\right)
$ is the signature resulting from reordering the one-electron creators into the
same order as the given Slater determinant.

In the CC wavefunction, all of the excitation operators commute with one
another.
Accounting for all possible orderings of the operators results in a permanent of
the parameters with identical rows, i.e.
\begin{equation*}
  \begin{vmatrix}
    t_{\mathbf{i}_1}^{\mathbf{a}_1} & \dots & t_{\mathbf{i}_N}^{\mathbf{a}_N}\\
    \vdots & \ddots & \vdots\\
    t_{\mathbf{i}_1}^{\mathbf{a}_1} & \dots & t_{\mathbf{i}_N}^{\mathbf{a}_N}
  \end{vmatrix}^+\\
\end{equation*}
Similarly, if all of the creation operators commute with one another, i.e.  they
are all even-electron creators, then
\begin{equation}
  \hspace{-3em}
  \ket{\Psi}
  =
  \sum_{
    \mathbf{m} \in \setsd
  }
  \left(
    \mathlarger{\sum}\limits_{
      \begin{smallmatrix}
        \\[0.2em]
        \{
        \mathbf{b}_1 \dots \mathbf{b}_{n_b}
        \}
        \subseteq \tilde{S}_{\mathbf{b}}
        \\
        \mathrm{sgn}
        \hat{A}^\dagger_{\mathbf{b}_1} \dots \hat{A}^\dagger_{\mathbf{b}_{n_b}}
        \ket{0}
        = \ket{\mathbf{m}}
      \end{smallmatrix}
    }
    \hspace{-2.8em}
    \mathrm{sgn}
    \left(
      \sigma(
      \hat{A}^\dagger_{\mathbf{b}_1} \dots \hat{A}^\dagger_{\mathbf{b}_{n_b}}
      )
    \right)
    \prod\limits_{i=1}^{n_b} C_{\mathbf{b}_i}
  \right)
  \ket{\mathbf{m}}
\end{equation}
where
\begin{equation*}
  \setsd = \left\{ \prod_{\hat{A}^\dagger_k \in T} \hat{A}^\dagger_k \ket{0} \middle| T \subseteq \tilde{S}_{\mathbf{b}} \right\}
\end{equation*}

For systems with an odd number of electrons, there must be at least one
odd-electron creator.
The anticommutation between these creators results in a determinant.
Unlike the permanent, the determinant of a matrix with identical rows is zero.
Therefore, there must be one odd-electron creator within a set of creation
operators that construct $\mathbf{m}$.
\begin{equation}
  \label{eq:FANCI_cc_quasiparticle}
  \hspace{-3em}
  \ket{\Psi}
  =
  \sum_{\mathbf{m} \in S_{\mathbf{m}}}
  \left(
    \mathlarger{\sum}\limits_{
      \begin{smallmatrix}
        \\[0.2em]
        \{
        \mathbf{b}_1 \dots \mathbf{b}_{n_b}
        \} \subseteq \tilde{S}_{\mathbf{b}},\;
        \mathbf{f} \in \tilde{S}_{\mathbf{f}}\\
        \mathrm{sgn}
        \hat{A}^\dagger_{\mathbf{b}_1} \dots \hat{A}^\dagger_{\mathbf{b}_{n_b}}
        \hat{A}^\dagger_{\mathbf{f}}
        \ket{0}
        = \ket{\mathbf{m}}
      \end{smallmatrix}
    }
    \hspace{-3.5em}
    \mathrm{sgn}
    \left(
      \sigma(
      \hat{A}^\dagger_{\mathbf{b}_1} \dots \hat{A}^\dagger_{\mathbf{b}_{n_b}}
      \hat{A}^\dagger_{\mathbf{f}}
      )
    \right)
    \left(
      \prod\limits_{i=1}^{n_b} C_{\mathbf{b}_i}
    \right)
    C_{\mathbf{f}}
  \right)
  \ket{\mathbf{m}}\\
\end{equation}
The derivation is provided in the Appendix~\ref{sec:appendix_cc_creation}.

In the case where only two-electron creation operators are used,
this wavefunction reduces to the Antisymmetrized Geminal Power (AGP)\cite{agp},
HF-Bogoliubov\cite{hfb_?}, or the BCS superconducting\cite{bcs} wavefunction
(Equation~\ref{eq:agp}).
\begin{equation}
  \label{eq:agp}
  \begin{split}
    \ket{\Psi} &= \exp\left( \sum_{ij} c_{ij} a_i^\dagger a_j^\dagger \right) \ket{0}\\
    &=
    \sum_{
      \mathbf{m} \in \setsd
    }
    \left(
      \mathlarger{\mathlarger{\sum}}_{
        \begin{smallmatrix}
          \{\mathbf{m}_1 \dots \mathbf{m}_{N/2}\} \subseteq \tilde{S}\\
          \mathrm{sgn} A^\dagger_{\mathbf{m}_1} \dots A^\dagger_{\mathbf{m}_{N/2}} \ket{0} = \ket{\mathbf{m}}\\
        \end{smallmatrix}
      }
      \hspace{-2em}
      \mathrm{sgn}(\sigma_{\mathbf{m}_1 \dots \mathbf{m}_{N/2}})
      \prod_{k=1}^{N/2} C_{\mathbf{m}_k}
    \right) \ket{\mathbf{m}}\\
  \end{split}
\end{equation}

Notice that this type of coupled-cluster wavefunction is \textit{not} size consistent and that it breaks particle number symmetry (unless it is restored with a projection onto correct particle number).

\subsection{TPS Variants}
\label{sec:tps_variants}
\subsubsection{TPS with Quasiparticles}
In the TPS wavefunction (Equation~\ref{eq:FANCI_tps_m}), each parameter in
$(M_k)^{n_k}$ describes the correlation between a spatial
orbital, $k$, and all of the other orbitals.
In the APG wavefunction (Equation~\ref{eq:apg}) and CC-motivated quasiparticle
wavefunction (Equation~\ref{eq:FANCI_cc_quasiparticle}), each
parameter is associated with a cluster of spatial orbitals (quasiparticle).
Then, we should be able to build a TPS-like wavefunction using creation
operators of an arbitrary number of electrons, rather than the one-electron
creation operators.
\begin{equation}
  \label{eq:FANCI_tps_quasiparticle}
    \ket{\Psi}
    =
    \sum_{
      \mathbf{m} \in \setsd
    }
    \left(
      \mathlarger{\mathlarger{\sum}}_{
        \begin{smallmatrix}
          \{\mathbf{m}_1 \dots \mathbf{m}_{n}\} \subseteq \tilde{S}\\
          \mathrm{sgn} A^\dagger_{\mathbf{m}_1} \dots A^\dagger_{\mathbf{m}_n} \ket{0} = \ket{\mathbf{m}}\\
        \end{smallmatrix}
      }
      \mathrm{sgn}(\sigma_{\mathbf{m}_1 \dots \mathbf{m}_n})
      \bigodot_{\mathbf{k} \in \tilde{S}}
      (M_{\mathbf{k}})^{\delta(\mathbf{k}, \{\mathbf{m}_1 \dots \mathbf{m}_n\})}
    \right)
    \ket{\mathbf{m}}\\
\end{equation}
where
\begin{equation*}
  \setsd =
  \left\{
    \prod_{\hat{A}^\dagger_{\mathbf{k}} \in T} \hat{A}^\dagger_{\mathbf{k}} \ket{0}
    \middle|
    T \subseteq \tilde{S}
  \right\}
\end{equation*}
$M_{\mathbf{k}}$ is a tensor that corresponds to the creation operator
$\hat{A}^\dagger_\mathbf{k}$, and $\delta$ is a function that checks if the
creator $\hat{A}^\dagger_\mathbf{k}$ is in a set of creators,
$\{\mathbf{m}_1 \dots \mathbf{m}_n\}$.
\begin{equation*}
  \delta(\mathbf{k}, \{\mathbf{m}_1 \dots \mathbf{m}_n\}) =
  \begin{cases}
    1 & \mbox{ if } \mathbf{k} \in \{\mathbf{m}_1 \dots \mathbf{m}_n\}\\
    0 & \mbox{ if } \mathbf{k} \not\in \{\mathbf{m}_1 \dots \mathbf{m}_n\}
  \end{cases}
\end{equation*}
Then, $(M_{\mathbf{k}})^0$ and $(M_{\mathbf{k}})^1$ are tensors that correspond
to the absence and presence of the creator $\hat{A}^\dagger_{\mathbf{k}}$,
respectively.

\subsubsection{TPS with Excitation Operators}
Just as the CC wavefunction can be rebuilt with creation operators, we can
rebuild the TPS wavefunction with excitation operators.
Each tensor, $t_{\hat{E}_\mathbf{i}^\mathbf{a}}$, can be associated with
$\hat{E}_{\mathbf{i}}^{\mathbf{a}}$, and has auxiliary indices that describe
the correlation between operators.
\begin{equation}
  \label{eq:FANCI_tps_excitation}
  \ket{\Psi} =
  \mathlarger{\sum}_{
    \hat{E}_{\mathbf{i}}^{\mathbf{a}} \in \setexc
  }
  \left(
    \mathlarger{\sum}_{
      \begin{smallmatrix}
        \{\hat{E}_{\mathbf{i}_1}^{\mathbf{a}_1} \dots \hat{E}_{\mathbf{i}_n}^{\mathbf{a}_n}\}
        \subseteq \tilde{S}_{\hat{E}}\\
        \mathrm{sgn} \prod_{k=1}^n \hat{E}_{\mathbf{i}_k}^{\mathbf{a}_k} = \hat{E}_{\mathbf{i}}^{\mathbf{a}}
      \end{smallmatrix}
    }
    \mathrm{sgn}(\sigma_{\hat{E}_{\mathbf{i}_1}^{\mathbf{a}_1} \dots \hat{E}_{\mathbf{i}_n}^{\mathbf{a}_n}})
    \bigodot_{\hat{E}_k \in \tilde{S}_{\hat{E}}}
    (t_{\hat{E}_{k}})^{
      \delta(
      \hat{E}_{k},
      \{\hat{E}_{\mathbf{i}_1}^{\mathbf{a}_1} \dots \hat{E}_{\mathbf{i}_n}^{\mathbf{a}_n}\}
      )
    }
  \right) \hat{E}_{\mathbf{i}}^{\mathbf{a}} \ket{\Phi_{\mathrm{HF}}}
\end{equation}
where
\begin{equation*}
  \setexc = \left\{ \prod_{\hat{E}_k \in T} \hat{E}_k \middle| T \subseteq \tilde{S}_{\hat{E}} \right\}
\end{equation*}
and $\delta$ describes the presence of an excitation operator,
$\hat{E}_{k}$, in the given set,
$\{\hat{E}_{\mathbf{i}_1}^{\mathbf{a}_1} \dots \hat{E}_{\mathbf{i}_n}^{\mathbf{a}_n}\}$
\begin{equation}
  \delta(
  \hat{E}_{k},
  \{\hat{E}_{\mathbf{i}_1}^{\mathbf{a}_1} \dots \hat{E}_{\mathbf{i}_n}^{\mathbf{a}_n}\}
  ) =
  \begin{cases}
    1 & \mbox{ if } \hat{E}_{k} \in
    \{\hat{E}_{\mathbf{i}_1}^{\mathbf{a}_1} \dots \hat{E}_{\mathbf{i}_n}^{\mathbf{a}_n}\}\\
    0 & \mbox{ if } \hat{E}_{k} \not\in
    \{\hat{E}_{\mathbf{i}_1}^{\mathbf{a}_1} \dots \hat{E}_{\mathbf{i}_n}^{\mathbf{a}_n}\}
  \end{cases}
\end{equation}

Comparing Equation~\ref{eq:FANCI_tps_excitation} with
Equation~\ref{eq:FANCI_cc},
the CC wavefunction can be considered a special case of this wavefunction, where
the tensor $(t_{\hat{E}_{k}})^{\delta_k}$ is $1$ if $\delta_{k} = 0$ and a
variable scalar value if $\delta_{k} = 1$.

\subsection{APG Generalized to Excitation Operators}
Just as the TPS wavefunction can be built with excitation operators
(Equation~\ref{eq:FANCI_tps_excitation}), we can rewrite the APG wavefunction
with excitation operators.
\begin{equation}
  \label{eq:apg_variant}
  \ket{\Psi} = \prod_{p=1}^n
  \left(
    \sum_{\hat{E}_{k} \in \tilde{S}_{\hat{E}}}
    t_{p;\hat{E}_{k}} \hat{E}_{k}
  \right) \ket{\Phi_{\mathrm{HF}}}
\end{equation}
where $n$ is the number of excitation operators that will be multiplied
together.
Unlike the CC wavefunction, which can generate all possible combinations of the
excitation operators, this wavefunction can only account for the combinations of
$n$ excitation operators.
The corresponding weight function in the FANCI notation is
\begin{equation}
  \label{eq:FANCI_apg_excitation}
  \ket{\Psi} =
  \mathlarger{\sum}_{
    \hat{E}_{\mathbf{i}}^{\mathbf{a}} \in \setexc
  }
  \left(
    \mathlarger{\sum}_{
      \begin{smallmatrix}
        \{\hat{E}_{\mathbf{i}_1}^{\mathbf{a}_1} \dots \hat{E}_{\mathbf{i}_n}^{\mathbf{a}_n}\}
        \subseteq \tilde{S}_{\hat{E}}\\
        \mathrm{sgn} \prod_{k=1}^n \hat{E}_k = \hat{E}_{\mathbf{i}}^{\mathbf{a}}
      \end{smallmatrix}
    }
    \mathrm{sgn}(\sigma_{\hat{E}_{\mathbf{i}_1}^{\mathbf{a}_1} \dots \hat{E}_{\mathbf{i}_n}^{\mathbf{a}_n}})
    \begin{vmatrix}
      t_{1;\hat{E}_{\mathbf{i}_1}^{\mathbf{a}_1}} & \dots & t_{1;\hat{E}_{\mathbf{i}_n}^{\mathbf{a}_n}}\\
      \vdots & \ddots  & \vdots\\
      t_{n;\hat{E}_{\mathbf{i}_1}^{\mathbf{a}_1}} & \dots & t_{n;\hat{E}_{\mathbf{i}_n}^{\mathbf{a}_n}}\\
    \end{vmatrix}^+
  \right) \hat{E}_{\mathbf{i}}^{\mathbf{a}} \ket{\Phi_{\mathrm{HF}}}
\end{equation}
where
\begin{equation*}
  \setexc =
  \left\{ \prod_{\hat{E}_k \in T} \hat{E}_k \middle| T \subseteq \tilde{S}_{\hat{E}} \right\}
\end{equation*}
and
$\mathrm{sgn}(\sigma_{\hat{E}_{\mathbf{i}_1}^{\mathbf{a}_1} \dots \hat{E}_{\mathbf{i}_n}^{\mathbf{a}_n}})$
is the signature of the permutation of the one-electron creation and
annihilation operators from the product of excitation operators,
$\hat{E}_{\mathbf{i}_1}^{\mathbf{a}_1} \cdots \hat{E}_{\mathbf{i}_n}^{\mathbf{a}_n}$,
to the given excitation operator, i.e.
\begin{equation}
  \hat{E}_{\mathbf{i}}^{\mathbf{a}} =
  \mathrm{sgn}(\sigma_{\hat{E}_{\mathbf{i}_1}^{\mathbf{a}_1} \dots \hat{E}_{\mathbf{i}_n}^{\mathbf{a}_n}})
  \hat{E}_{\mathbf{i}_1}^{\mathbf{a}_1} \cdots \hat{E}_{\mathbf{i}_n}^{\mathbf{a}_n}
\end{equation}

Other combinations of excitation operators can be included into
Equation~\ref{eq:apg_variant} via the summation.
\begin{equation}
  \label{eq:apg_variant_2}
  \ket{\Psi} =
  \sum_{n \in P} \frac{1}{n!}
  \prod_{p=1}^n
  \left(
    \sum_{\hat{E}_{k} \in \tilde{S}_{\hat{E}}}
    t_{p;\hat{E}_{k}} \hat{E}_{k}
  \right) \ket{\Phi_{\mathrm{HF}}}
\end{equation}
where $P$ is the allowed number of excitation operators that can be combined.
Recall that removing an index (so that all the geminals are identical)
in the traditional APG wavefunction form leads to the AGP wavefunction.
Similarly, when one removes the index, $p$, from the excitation-based APG
wavefunction form and allows all possible combinations of excitation operators,
one obtains the CC wavefunction:
\begin{equation}
  \begin{split}
    \ket{\Psi_{\mathrm{CC}}}
    &= \sum_{n=0}^\infty \frac{1}{n!}
    \prod_{p=1}^n
    \left(
      \sum_{\hat{E}_{k} \in \tilde{S}_{\hat{E}}}
      t_{\hat{E}_{k}} \hat{E}_{k}
    \right) \ket{\Phi_{\mathrm{HF}}}\\
    &= \sum_{n=0}^\infty \frac{1}{n!}
    \left(
      \sum_{\hat{E}_{k} \in \tilde{S}_{\hat{E}}}
      t_{\hat{E}_{k}} \hat{E}_{k}
    \right)^n \ket{\Phi_{\mathrm{HF}}}\\
    &= \exp
    \left(
      \sum_{\hat{E}_{k} \in \tilde{S}_{\hat{E}}}
      t_{\hat{E}_{k}} \hat{E}_{k}
    \right) \ket{\Phi_{\mathrm{HF}}}\\
  \end{split}
\end{equation}

\subsection{General Quasiparticle Wavefunctions}
In the Equation~\ref{eq:FANCI_cc_quasiparticle}, any creation operator can be
used (within the sets $\tilde{S}_{\mathbf{b}}$ and $\tilde{S}_{\mathbf{f}}$) to construct a Slater determinant.
Similarly, we can generalize the APG wavefunction to include all even-electron
creation operators.
\begin{equation}
  \label{eq:quasiparticle}
  \ket{\Psi} =
  \prod_{p=1}^{n}
  \left(
    \sum_{\hat{A}^\dagger_k \in \tilde{S}_{\mathbf{b}}} C_{p;\mathbf{m}_k} \hat{A}^\dagger_{k}
  \right)
  \ket{0}
\end{equation}
where $n$ is the number of quasiparticles in the wavefunction (i.e. number of
operators used to construct a Slater determinant), and $\tilde{S}_{\mathbf{b}}$ is a set of
even-electron creation operators.
In the FANCI formulation,
\begin{equation}
  \label{eq:FANCI_boson}
  \ket{\Psi}
  =
  \sum_{
    \mathbf{m} \in \setsd
  }
  \left(
    \mathlarger{\mathlarger{\sum}}_{
      \begin{smallmatrix}
        \{\mathbf{m}_1 \dots \mathbf{m}_{n}\}\subseteq \tilde{S}_{\mathbf{b}}\\
        \mathrm{sgn} A^\dagger_{\mathbf{m}_1} \dots A^\dagger_{\mathbf{m}_n} \ket{0} = \ket{\mathbf{m}}\\
      \end{smallmatrix}
    }
    \mathrm{sgn}(\sigma_{\mathbf{m}_1 \dots \mathbf{m}_n})
    \begin{vmatrix}
      C_{1;\mathbf{m}_1} &\dots & C_{1;\mathbf{m}_{n}}\\
      \vdots & \ddots & \vdots\\
      C_{n;\mathbf{m}_1} &\dots & C_{n;\mathbf{m}_{n}}\\
    \end{vmatrix}^+
  \right)
  \ket{\mathbf{m}}
\end{equation}
where $\tilde{S}_{\mathbf{b}}$ is a set of allowed even-electron creation operators,
\begin{equation*}
  \setsd = \left\{ \prod_{\hat{A}^\dagger_k \in T} \hat{A}^\dagger_k \ket{0} \middle| T \subseteq \tilde{S}_{\mathbf{b}} \right\}
\end{equation*}
and $\mathrm{sgn}(\sigma_{\mathbf{m}_1 \dots \mathbf{m}_n})$ is the signature of the
permutation needed to reorder the one-electron creation operators from the
product of quasiparticle creation operators to the given Slater determinant.
\begin{equation}
  \prod_{i \in \mathbf{m}} a^\dagger_i =
  \mathrm{sgn}(\sigma_{\mathbf{m}_1 \dots \mathbf{m}_n})
  \prod_{k=1}^n \prod_{i \in \mathbf{m}_k} a^\dagger_i
\end{equation}
Note that the zero-electron creation operator can be considered as an even-electron
creation operator.

Similarly, the wavefunction constructed using only odd-electron creation
operator can be expressed using a determinant.
\begin{equation}
  \label{eq:FANCI_fermion}
  \ket{\Psi}
  =
  \sum_{
    \mathbf{m} \in \setsd
  }
  \left(
    \mathlarger{\mathlarger{\sum}}_{
      \begin{smallmatrix}
        \{\mathbf{m}_1 \dots \mathbf{m}_{n}\}\subseteq \tilde{S}_{\mathbf{f}}\\
        \mathrm{sgn} A^\dagger_{\mathbf{m}_1} \dots A^\dagger_{\mathbf{m}_n} \ket{0} = \ket{\mathbf{m}}\\
      \end{smallmatrix}
    }
    \mathrm{sgn}(\sigma_{\mathbf{m}_1 \dots \mathbf{m}_n})
    \begin{vmatrix}
      C_{1;\mathbf{m}_1} &\dots & C_{1;\mathbf{m}_{n}}\\
      \vdots & \ddots & \vdots\\
      C_{n;\mathbf{m}_1} &\dots & C_{n;\mathbf{m}_{n}}\\
    \end{vmatrix}^-
  \right) \ket{\mathbf{m}}
\end{equation}
where
\begin{equation*}
  \setsd = \left\{ \prod_{\hat{A}^\dagger_k \in T} \hat{A}^\dagger_k \ket{0} \middle| T \subseteq \tilde{S}_{\mathbf{f}} \right\}
\end{equation*}
and $\tilde{S}_{\mathbf{f}}$ is a set of selected odd-electron creation operators.

When both even and odd-electron creation operators are present in $\tilde{S}$, the
interchange of creators commutes or anticommutes depending on the
pair of creators, and additional structure is necessary to account for this
behaviour.
First, we distinguish between the set of even and odd orbitals with
$\mathbf{b}_i$ and $\mathbf{f}_j$, respectively.
A given Slater determinant $\mathbf{m}$ is constructed with $n = n_b + n_f$
creation operators, where $n_b$ is the number of even-electron creators and $n_f$ is
the number of odd-electron creators.
A permanent is needed to account for all possible ordering of the even-electron
creators and the resulting commutations.
A determinant is needed to account for all possible ordering of the
odd-electron creators and the resulting anticommutations.
Finally, the commutation between an even-electron and an odd-electron creator
can be accounted for by a sum over all possible selections of the $n_b$
even-electron creators from $n$ positions, i.e.
$\{i^b_1 \dots i^b_{n_b}\} \subseteq
\{\mathbf{b}_1 \dots \mathbf{b}_{n_b} \mathbf{f}_1 \dots \mathbf{f}_{n_f} \}$.
The positions of the odd-electron creators are the remaining $n_f$ positions,
i.e.
$\{i^f_1 \dots i^f_{n_f}\} =
\{\mathbf{b}_1 \dots \mathbf{b}_{n_b} \mathbf{f}_1 \dots \mathbf{f}_{n_f} \}
\setminus \{i^b_1 \dots i^b_{n_b}\}$.
Then, we can construct the generalized quasiparticle wavefunction
(Equation~\ref{eq:quasiparticle}) in which any set of creators can be used.
\begin{equation}
  \label{eq:FANCI_quasiparticle}
  \hspace{-6em}
  \ket{\Psi}
  =
  \sum_{\mathbf{m}}
  \left(
    \mathlarger{\sum}_{
      \begin{smallmatrix}
        \\[0.2em]
        \{
        \mathbf{b}_1 \dots \mathbf{b}_{n_b}
        \} \subseteq \tilde{S}_{\mathbf{b}},\;
        \{
        \mathbf{f}_1 \dots \mathbf{f}_{n_f}
        \} \subseteq \tilde{S}_{\mathbf{f}}\\
        \mathrm{sgn}
        \hat{A}^\dagger_{\mathbf{b}_1} \dots \hat{A}^\dagger_{\mathbf{b}_{n_b}}
        \hat{A}^\dagger_{\mathbf{f}_1} \dots \hat{A}^\dagger_{\mathbf{f}_{n_f}}
        \ket{0}
        = \ket{\mathbf{m}}
      \end{smallmatrix}
    }
    \hspace{-4em}
    \mathrm{sgn}
    (\sigma_{
      \hat{A}^\dagger_{\mathbf{b}_1} \dots \hat{A}^\dagger_{\mathbf{b}_{n_b}}
      \hat{A}^\dagger_{\mathbf{f}_1} \dots \hat{A}^\dagger_{\mathbf{f}_{n_f}}
    })
    \hspace{-5.2em}
    \mathlarger{\sum}_{
      \begin{smallmatrix}
        \\[0.2em]
        \{i^b_1 \dots i^b_{n_b}\} \subseteq
        \{\mathbf{b}_1 \dots \mathbf{b}_{n_b}, \mathbf{f}_1 \dots \mathbf{f}_{n_f} \}\\
        \{i^f_1 \dots i^f_{n_f}\} =
        \{\mathbf{b}_1 \dots \mathbf{b}_{n_b}, \mathbf{f}_1 \dots \mathbf{f}_{n_f} \}
        \setminus \{i^b_1 \dots i^b_{n_b}\}\\
      \end{smallmatrix}
    }
    \begin{vmatrix}
      C^b_{1 i^b_1} & \hspace{-0.6em}\dots & \hspace{-0.6em}C^b_{1 i^b_{n_b}}\\[-0.2em]
      \vdots & \hspace{-0.6em}\ddots & \hspace{-0.6em}\vdots\\[-0.6em]
      C^b_{n_b i^b_1} & \hspace{-0.6em}\dots & \hspace{-0.6em}C^b_{n_b i^b_{n_b}}\\
    \end{vmatrix}^+
    \begin{vmatrix}
      C^f_{1 i^f_1} & \hspace{-0.6em}\dots & \hspace{-0.4em}C^f_{1 i^f_{n_f}}\\[-0.2em]
      \vdots & \hspace{-0.6em}\ddots & \hspace{-0.4em}\vdots\\[-0.6em]
      C^f_{n_f i^f_1} & \hspace{-0.6em}\dots & \hspace{-0.4em}C^f_{n_f i^f_{n_f}}\\
    \end{vmatrix}^-
  \right)
  \ket{\mathbf{m}}\\
\end{equation}
  where
\begin{equation*}
  \setsd = \left\{ \prod_{\hat{A}^\dagger_k \in T} \hat{A}^\dagger_k \ket{0} \middle| T \subseteq \tilde{S}_{\mathbf{b}} \cup \tilde{S}_{\mathbf{f}} \right\}
\end{equation*}
\begin{equation*}
  \prod_{i \in \mathbf{m}} a^\dagger_i =
    \mathrm{sgn}
    (\sigma_{
      \hat{A}^\dagger_{\mathbf{b}_1} \dots \hat{A}^\dagger_{\mathbf{b}_{n_b}}
      \hat{A}^\dagger_{\mathbf{f}_1} \dots \hat{A}^\dagger_{\mathbf{f}_{n_f}}
    })
    \prod_{k=1}^{n_b} A^\dagger_{\mathbf{b}_k}
    \prod_{l=1}^{n_f} A^\dagger_{\mathbf{f}_l}
\end{equation*}
and $C$ is a $n\times\dim(\tilde{S}_{\mathbf{b}} \cup \tilde{S}_{\mathbf{f}})$
matrix.
Note that the set of orbitals, $\mathbf{b}_i$ and $\mathbf{f}_j$, correspond to
a column in $C$ and thus are used as a column index.
Here, the column indices are made explicit using the index $i$.

The derivation and more details are in the Appendix~\ref{sec:appendix_quasiparticle}.
% TODO: mention that all geminal flavors in 2017 paper are FANCI
% TODO: mention product/quotient of determinants
% TODO: mention CI of FANCI is universal

\subsection{Changing Solvers}
Often times, the algorithm for optimizing the parameters is synonymous with the
wavefuncton ansatz.
For example, DMRG is often associated with MPS wavefunctions, and Quantum
Monte-Carlo (QMC) is often associated with FCI wavefunctions.
Using different algorithms to optimize the parameters will change the cost and
the reliability of the wavefunction.
Certain algorithms can be applied to a wide range of wavefunction structures and
systems;
whereas specialized algorithms are cheaper, but are often limited to specific
wavefunction structures and systems. % (FIXME)

For example, the DMRG algorithm with the MPS wavefunction provides variational
results and is effective in describing linear systems.
The algorithm is specific to the MPS wavefunction, wherein the orbitals are
ordered such that adjacent orbitals are more correlated than the rest.
As a result, DMRG must be extended beyond MPS to more general TPS forms, but
these algorithms do not seem to be as elegant or as computationally efficient.
Solving the TPS wavefunction as a projected Schr\"{o}dinger equation
mitigates certain complications that are present in DMRG, such as ordering of the
orbitals and generalizing a one-dimensional algorithm to multiple dimensions. It
is true, however, that the tensor structures that are most efficiently optimized
in a variational ansatz are, at least in all cases we have considered, the same
as those that are most efficiently optimized using the projected Schr\"{o}dinger
equation.

\subsection{Generalization}
Above, we constructed new wavefunction structures by constructing the CC
wavefunction with creation operators and the TPS and APG wavefunctions using
excitation operators.
Effectively, the parameters are changed from the contributions of creation
operators to those of excitation operators, and vice versa.
Since both operators originate from orbitals, the wavefunctions are
size-consistent if $f$ is selected appropriately, provided the orbitals are
localized.
In Equation~\ref{eq:FANCI_cc_quasiparticle}, \ref{eq:FANCI_boson},
\ref{eq:FANCI_fermion}, and \ref{eq:FANCI_quasiparticle}, particle number
symmetry is not conserved, which is not a problem since the wavefunction can be
optimized and projected onto the appropriate particle number.
Then, just as excitation operators change the state of a reference Slater
determinant, products of an arbitrary number of creation and annihilation
operators can be used to explore different particle numbers with respect to the
reference.
For example, the product of creators and annihilators, where the number of
creators exceeds the number of annihilators, will ionize a Slater determinant.
Expressing the wavefunction with respect to different operators will result in
different sets of parameters, even if these operators make no reference to
the Slater determinants or creation operators.
For example, grid-based wavefunctions will have parameters for each point in
space.

So far, we have seen weight functions, $f$, in various forms:
the CC wavefunction (Equation~\ref{eq:FANCI_cc}) uses a cumulant;
the TPS wavefunction (Equation~\ref{eq:FANCI_tps}) uses a tensor product;
the even-electron quasiparticle wavefunction (Equation~\ref{eq:FANCI_fermion})
uses a permanent;
the odd-electron quasiparticle wavefunction (Equation~\ref{eq:FANCI_boson})
uses a determinant;
and the generalized quasiparticle wavefunction
(Equation~\ref{eq:FANCI_quasiparticle}) uses some mix of a permanent and
determinant (an immanent).
In Equations~\ref{eq:FANCI_apg}, \ref{eq:FANCI_cc_quasiparticle},
\ref{eq:FANCI_tps_quasiparticle}, and \ref{eq:FANCI_quasiparticle},
creators of arbitrary number of electrons are combined in fairly complicated
manner.
Since creators have distinct commutative (and anticommutative) relations with one
another, the corresponding $f$ must be symmetric (and antisymmetric) with
respect to interchange of different creation operators (or parameters).
Grouping together all the terms that correspond to the same set of creation
operators results in a sum over all combinations of products of parameters.
These combinatorial variants of product functions seem to be useful for
size-consistency since they are multiplicatively separable by construction.
Therefore, we can construct novel wavefunctions with quasiparticle origins using
different symmetric (or antisymmetric) polynomial functions, including but
not limited to determinant, permanent, immanent\cite{immanant}, pfaffian
\cite{pfaffian}, hafnian\cite{hafnian, hafnian_borchardt}, hyperdeterminant
\cite{hyperdeterminant}, multidimensional permanent\cite{multi_permanent},
hyperpfaffian\cite{hyperpfaffian}, hyperhafnian\cite{hyperhafnian},
and mixed discriminant\cite{mixed_discriminant}.
Unfortunately, among all the aforementioned size-consistent combinatoric forms,
only the determinant can be evaluated in polynomial time.
If we disregard size-consistency and do not require any quasiparticle structure,
then any function can be chosen.
One alternative, however, is to use a ratio of a product of determinants as the
overlap; this form retains size-consistency and computational feasibility, and
generalizes a single Slater determinant (one determinant in the numerator, none
in the denominator) and the APr2G wavefunction (a special case with one
structure determinant in the denominator and the determinant of its element-wise
square in the numerator).
If the total number of determinants is odd, this wavefunction describes a normal
fermionic wavefunction, while if the number of determinants is even, this
wavefunction represents a bosonic seniority-zero structure.

A great deal of flexibility is available within the proposed wavefunction
framework:
the set of Slater determinants, $S$, can be any set of orthonormal (for
convenience) Slater determinants;
the parameters, $\vec{P}$, can be any set of numbers that describe
the wavefunction or the operators with which it is built;
the parameterizing function, $f$, can be any function that maps the parameters
to a coefficient for a given Slater determinant.
We hope to find the combination that effectively models the optimized
coefficients of the FCI wavefunction (for accuracy) using the minimal number of
parameters (for cost) for as many systems as possible (for generality).

\section{Conclusion}
\label{sec:fanci_conclusion}
In the proposed framework, any multideterminant wavefunction can be expressed
with respect to the components $S$, $\vec{P}$, and $f$.
A wavefunction can be characterized using only these components and different
combinations will result in different wavefunction characteristics.
Then, we can systematically develop new structures by simply finding novel
combinations.
While the proposed wavefunctions are not necessarily cheap to compute, the
general ansatz does not need to be cheap to be effective.
For example, the CC wavefunction with all orders of excitations, the TPS
wavefunction with infinitely large tensors, and the quasiparticle wavefunctions
with $N$ electron quasiparticles are no less expensive than the FCI
wavefunction.
However, different wavefunction structures inspire different approximations and
new algorithms that reduce the computational cost to a tractable level.
The greatest advantage to this perspective is that it is pragmatic:
approximations and algorithms that were restricted to one wavefunction can
be generalized to others using the FANCI framework;
and many different methods can be implemented computationally using a common
framework.
Though we will defer detailed discussions on dynamical correlation corrections
and orbital optimization to future papers, these elaborations to FANCI are
clearly possible and are performed using similar techniques to what one would
use in traditional selected CI and CC methods.
Many of the proposed methods have been implemented in an open-source library called \texttt{Fanpy}, which will be presented elsewhere.

\section{Acknowledgements}
The authors thank NSERC, Compute Canada, McMaster University, and the Canada
Research Chairs for funding.
RAMQ acknowledges financial support from the University of Florida in the form
of a start-up grant.

\newpage
\bibliography{references}

\section{Appendix}
\subsection{HF}
\label{sec:appendix_hf}
\begin{equation}
  \begin{split}
    \ket{\Psi_{\mathrm{HF}}}
    &= \prod_{i=1}^N \left( \sum_{j=1}^{2K} a^\dagger_j U_{ji} \right) \ket{0}\\
    &= \left( \sum_{j_1=1}^{2K} a^\dagger_{j_1} U_{j_11} \right)
    \left( \sum_{j_2=1}^{2K} a^\dagger_{j_2} U_{j_22} \right)
    \dots
    \left( \sum_{j_N=1}^{2K} a^\dagger_{j_N} U_{j_NN} \right) \ket{0}\\
    &=
    \sum_{j_1=1}^{2K} \sum_{j_2=1}^{2K} \dots \sum_{j_N=1}^{2K}
    a^\dagger_{j_1} a^\dagger_{j_2} \dots a^\dagger_{j_N}
    U_{j_11} U_{j_22} \dots U_{j_NN} \ket{0}\\
  \end{split}
\end{equation}
We can group together the terms that result in the same Slater determinant,
defined here by the set of creation operators,
$\{a^\dagger_{j_1} a^\dagger_{j_2} \dots a^\dagger_{j_N}\}$.
Since the order of operators only affects the sign of the Slater determinant, we
can split the sum over all indices into a sum over the set of creation
operators and a sum over the different orderings of the given set of creation
operators.
\begin{equation}
  \begin{split}
    \ket{\Psi_{\mathrm{HF}}}
    &=
    \sum_{j_1 < j_2 \dots < j_N}^{2K}
    \sum_{\sigma \in S_N} U_{j_{\sigma(1)}1} U_{j_{\sigma(2)}2} \dots U_{j_{\sigma(N)}N}
    a^\dagger_{j_{\sigma(1)}} a^\dagger_{j_{\sigma(2)}} \dots a^\dagger_{j_{\sigma(N)}}
    \ket{0}\\
    &=
    \sum_{j_1 < j_2 \dots < j_N}^{2K}
    \sum_{\sigma \in S_N} \mathrm{sgn}(\sigma) U_{j_{\sigma(1)}1} U_{j_{\sigma(2)}2} \dots U_{j_{\sigma(N)}N}
    a^\dagger_{j_1} a^\dagger_{j_2} \dots a^\dagger_{j_N}
    \ket{0}\\
  \end{split}
\end{equation}
where $\mathrm{sgn}(\sigma)$ is the signature of the permutation $\sigma$.
\begin{equation}
  \mathrm{sgn}(\sigma) =
  \begin{cases}
    1 &\mbox{ if $\sigma$ is even}\\
    -1 &\mbox{ if $\sigma$ is odd}\\
  \end{cases}
\end{equation}
Using the definition of the determinant,
\begin{equation}
  \begin{split}
    | A |^-
    &= \sum_{\sigma \in S_N} \mathrm{sgn}(\sigma) \prod_{i=1}^N A_{i \sigma(i)}\\
    &= \sum_{\sigma \in S_N} \mathrm{sgn}(\sigma) \prod_{i=1}^N A_{\sigma(i) i}\\
  \end{split}
\end{equation}
we find
\begin{equation}
  \begin{split}
    \ket{\Psi_{\mathrm{HF}}}
    &=
    \sum_{j_1 < j_2 \dots < j_N}^K
    \begin{vmatrix}
      U_{j_11} & U_{j_12} & \dots & U_{j_1N}\\
      U_{j_21} & U_{j_22} & \dots & U_{j_2N}\\
      \vdots & \vdots & \ddots & \vdots\\
      U_{j_N1} & U_{j_N2} & \dots & U_{j_NN}\\
    \end{vmatrix}^-
    a^\dagger_{j_1} a^\dagger_{j_2} \dots a^\dagger_{j_N}
    \ket{0}\\
    &=
    \sum_{\mathbf{m}} |U(\mathbf{m})|^- \ket{\mathbf{m}}
  \end{split}
\end{equation}
where $\mathbf{m} = \{j_1, j_2, \dots, j_N\}$,
$\ket{\mathbf{m}} = a^\dagger_{j_1} a^\dagger_{j_2} \dots a^\dagger_{j_N}
\ket{0}$ where $j_1 < j_2 < \dots < j_N$,
and $U(\mathbf{m})$ is a submatrix of $U$ composed of rows that correspond to
$\mathbf{m}$.

\subsection{APG}
\label{sec:appendix_apg}
For convenience, let $P = \frac{N}{2}$.
\begin{equation*}
  \begin{split}
    \ket{\Psi_{\mathrm{APG}}}
    &= \prod_{p=1}^P \sum_{ij}^{2K} C_{p;ij} a_i^\dagger a_{j}^\dagger \ket{0}\\
    &= \left( \sum_{i_1 j_1}^{2K} a^\dagger_{i_1} a^\dagger_{j_1} C_{1;i_1 j_1} \right)
    \left( \sum_{i_2 j_2}^{2K} a^\dagger_{i_2} a^\dagger_{j_2} C_{2;i_2 j_2} \right)
    \dots
    \left( \sum_{i_P j_P}^{2K} a^\dagger_{i_P} a^\dagger_{j_P} C_{P;i_P j_P} \right) \ket{0}\\
    &=
    \sum_{i_1 j_1}^{2K} \sum_{i_2 j_2}^{2K} \dots \sum_{i_P j_P}^{2K}
    a^\dagger_{i_1} a^\dagger_{j_1} a^\dagger_{i_2} a^\dagger_{j_2} \dots a^\dagger_{i_P} a^\dagger_{j_P}
    C_{1;i_1 j_1} C_{2;i_2 j_2} \dots C_{P;i_P j_P} \ket{0}\\
  \end{split}
\end{equation*}
Just as in the HF derivation, we can group together the creation operators,
$\{a^\dagger_{i_1}, a^\dagger_{j_1}, \dots, a^\dagger_{i_P}, a^\dagger_{j_P}\}$, that
corresponds to the same Slater determinant.
However, there are multiple ways to construct a Slater determinant from a set of
\textit{pairs} of creators,
$\{a^\dagger_{i_1}a^\dagger_{j_1}, a^\dagger_{i_2}a^\dagger_{j_2}, \dots, a^\dagger_{i_P}a^\dagger_{j_P}\}$
(Note the placement of commas).
Since the Slater determinant is expressed with respect to a specific ordering of
the one-electron creation operators, the sign resulting from ordering the
product of creator pairs to this specific ordering must be taken
into account.
Then, the sum over all indices can be split into a sum over each Slater
determinant, a sum over different orbital pairs that construct the given Slater
determinant, and a sum over the permutation of the orbital pair creation operators.
\begin{equation}
  \begin{split}
    \ket{\Psi_{\mathrm{APG}}}
    &=
    \sum_{\mathbf{m} \in S}
    \mathlarger{\sum}_{
      \begin{smallmatrix}
        \{i_1j_1, \dots, i_Pj_P\} = \mathbf{m}
      \end{smallmatrix}
    }
    \sum_{\tau \in S_P}
    C_{1;i_{\tau(1)} j_{\tau(1)}} \dots C_{P;i_{\tau(P)} j_{\tau(P)}}
    a^\dagger_{i_{\tau(1)}} a^\dagger_{j_{\tau(1)}} \dots
    a^\dagger_{i_{\tau(P)}} a^\dagger_{j_{\tau(P)}}
    \ket{0}\\
    &=
    \sum_{\mathbf{m} \in S}
    \mathlarger{\sum}_{
      \begin{smallmatrix}
        \{i_1j_1, \dots, i_Pj_P\} = \mathbf{m}
      \end{smallmatrix}
    }
    \mathrm{sgn} \big(\sigma(i_1 j_1, \dots, i_P j_P)\big)
    \sum_{\tau \in S_P}
    C_{1;i_{\tau(1)} j_{\tau(1)}} \dots C_{P;i_{\tau(P)} j_{\tau(P)}}
    a^\dagger_{i_1} a^\dagger_{j_1} \dots a^\dagger_{i_P} a^\dagger_{j_P}
    \ket{0}\\
  \end{split}
\end{equation}
where $\sigma$ is the permutation for ordering the creation operators to that of
the Slater determinant and $\tau$ is the permutation of the given pairs of
creation operators.
Note that the pair of creation operators commute with one another.
From the definition of a permanent,
\begin{equation}
  \begin{split}
    | A |^+
    &= \sum_{\tau \in S_N} \prod_{i=1}^N A_{i \tau(i)}\\
    &= \sum_{\tau \in S_N} \prod_{i=1}^N A_{\tau(i) i}\\
  \end{split}
\end{equation}
we find
\begin{equation}
  \begin{split}
    \ket{\Psi_{\mathrm{APG}}}
    &=
    \sum_{\mathbf{m} \in S}
    \mathlarger{\sum}_{
      \begin{smallmatrix}
        \{i_1, j_1, \dots, i_P, j_P\} = \mathbf{m}
      \end{smallmatrix}
    }
    \mathrm{sgn} \big(\sigma(i_1, j_1, \dots, i_P, j_P)\big)
    \begin{vmatrix}
      C_{1;i_1 j_1} & \dots & C_{1; i_P j_P}\\
      \vdots & \ddots & \vdots\\
      C_{P;i_1 j_1} & \dots & C_{P; i_P j_P}\\
    \end{vmatrix}^+
    a^\dagger_{i_1} a^\dagger_{j_1} \dots a^\dagger_{i_P} a^\dagger_{j_P}
    \ket{0}\\
    &=
    \sum_{\mathbf{m} \in S}
    \mathlarger{\sum}_{
      \begin{smallmatrix}
        \{\mathbf{m}_1 \dots \mathbf{m}_P\}\\
        \mathrm{sgn} A^\dagger_{\mathbf{m}_1} \dots A^\dagger_{\mathbf{m}_P} \ket{0} = \ket{\mathbf{m}}
        % \bigcup_{p=1}^{P} \mathbf{m}_p = \mathbf{m}\\
        % \mathbf{m}_p \cap \mathbf{m}_q = 0 \; \forall p \neq q
      \end{smallmatrix}
    }
    \mathrm{sgn} \big(\sigma(\mathbf{m}_1 \dots \mathbf{m}_P)\big)
    |C(\mathbf{m}_1, \dots, \mathbf{m}_P)|^+
    \ket{\mathbf{m}}\\
  \end{split}
\end{equation}
where $\mathbf{m}_p = \{i_p, j_p\}$,
$A^\dagger_{\mathbf{m}_p} = a^\dagger_{i_p} a^\dagger_{j_p}$,
$\ket{\mathbf{m}} = a^\dagger_{i_1} a^\dagger_{j_1} \dots a^\dagger_{i_P}
a^\dagger_{j_P} \ket{0}$,
and $C(\mathbf{m}_1, \dots, \mathbf{m}_P)$ is a submatrix of $C$ composed of
columns that correspond to the orbital pairs, $\{\mathbf{m}_1 \dots \mathbf{m}_P\}$.

\subsection{Product of Linear Combinations of Operators}
Using the two examples above, we can generalize the approach to reformulating
wavefunctions of the form
\begin{equation}
  \ket{\Psi} = \prod_{i=1}^n \sum_j C_{ij} \hat{Q}_j \ket{\theta}
\end{equation}
where $\hat{Q}_{j}$ is an operator, $C_{ij}$ is a coefficient that corresponds
to the index $i$ and operator $\hat{Q}_j$, and $\theta$ is some
vacuum/reference.
Taking the same approach as above,
\begin{equation}
  \label{eq:approach}
  \begin{split}
    \ket{\Psi} &= \prod_{i=1}^n \sum_j C_{ij} \hat{Q}_j \ket{\theta}\\
    &= \left( \sum_{j_1} C_{1j_1} \hat{Q}_{j_1} \right) \dots
    \left( \sum_{j_n} C_{nj_n} \hat{Q}_{j_n} \right)
    \ket{\theta}\\
    &= \sum_{j_1}\dots \sum_{j_n} C_{1j_1} \dots C_{nj_n}
    \hat{Q}_{j_1} \dots \hat{Q}_{j_n} \ket{\theta}\\
    &=
    \underbrace{
      \hspace{2.2em}
      \sum_{\mathbf{m}}
      \hspace{2.2em}
    }_{sum over Slater determinants}
    \hspace{0.5em}
    \underbrace{
      \hspace{1em}
      \sum_{\{\hat{Q}_{j_1} \dots \hat{Q}_{j_n}\} \mapsto \mathbf{m}}
      \hspace{1em}
    }_{sum over all $\{\hat{Q}_{j_1} \dots \hat{Q}_{j_n}\}$ that produce the given Slater determinant}
    \hspace{0.5em}
    \underbrace{
      \hspace{1em}
      \mathrm{sgn}\big(\sigma(\hat{Q}_{j_1} \dots \hat{Q}_{j_n})\big)
      \hspace{1em}
    }_{signature for ordering the one-electron operators}
    \hspace{0.5em}
    \underbrace{
      \sum_{\tau \in S_n}
      \mathrm{sgn}(\tau)
      C_{1j_{\tau(1)}} \dots C_{nj_{\tau(n)}}
    }_{sum over the permutation of the given operators}
    \ket{\mathbf{m}}\\
  \end{split}
\end{equation}
Depending on the commutation/anticommutation relationships between the elements
of $\{\hat{Q}_{j_1} \dots \hat{Q}_{j_n}\}$, the $\mathrm{sgn}(\tau)$ is
different.

If all operators are expressed with an even number of one-electron
operators (which we will denote as even-electron operators), then
\begin{equation*}
  \mathrm{sgn}(\tau) = 1
\end{equation*}
and
\begin{equation}
  \sum_{\tau \in S_n}
  \mathrm{sgn}(\tau)
  C_{1j_{\tau(1)}} \dots C_{nj_{\tau(n)}}
  =
  \begin{vmatrix}
    C_{1j_1} & \dots & C_{1j_n}\\
    \vdots & \ddots & \vdots\\
    C_{nj_1} & \dots & C_{nj_n}\\
  \end{vmatrix}^+
\end{equation}

If all operators are expressed with an odd number of one-electron
operators (which we will denote as odd-electron operators), then
\begin{equation*}
  \mathrm{sgn}(\tau) =
  \begin{cases}
    1& \mbox{ if even}\\
    -1& \mbox{ if odd}\\
  \end{cases}
\end{equation*}
and
\begin{equation}
  \sum_{\tau \in S_n}
  \mathrm{sgn}(\tau)
  C_{1j_{\tau(1)}} \dots C_{nj_{\tau(n)}}
  =
  \begin{vmatrix}
    C_{1j_1} & \dots & C_{1j_n}\\
    \vdots & \ddots & \vdots\\
    C_{nj_1} & \dots & C_{nj_n}\\
  \end{vmatrix}^-
\end{equation}

If the operators are a mix of even and odd-electron operators, then $\tau$ must
be split into three components: $\tau_{b}$ for permutation of the positions of
even-electron operators, $\tau_{f}$ for permutation of the positions of
odd-electron operators, and $\tau_{bf}$ for remaining permutations that mix the
positions of even-electron operators with the odd-electron operators.
\begin{equation}
  \begin{split}
    \tau &= \tau_b \tau_f \tau_{bf}\\
    \mathrm{sgn}(\tau) &= \mathrm{sgn}(\tau_b) \mathrm{sgn}(\tau_f) \mathrm{sgn}(\tau_{bf})\\
    \mathrm{sgn}(\tau_b) &= 1\\
    \mathrm{sgn}(\tau_f) &=
    \begin{cases}
      1& \mbox{ if even}\\
      -1& \mbox{ if odd}\\
    \end{cases}\\
    \mathrm{sgn}(\tau_{bf}) &= 1\\
  \end{split}
\end{equation}
Let there be $n_b$ even-electron operators and  $n_f$ odd-electron operators.
\begin{equation*}
  n = n_b + n_f
\end{equation*}
We will assume (with no loss of generality) that the first $n_b$ columns of the
coefficients correspond to the even-electron and the rest to the odd-electron
operators.
Then $\tau_{bf}$ would correspond to swapping the first $n_b$ columns with
the rest of the columns.
After the swapped columns are obtained, the $\tau_b$ will permute the first
$n_b$ columns and $\tau_f$ will permute the others.
\begin{equation}
  \begin{split}
    &\sum_{\tau \in S_n}
    \mathrm{sgn}(\tau)
    C_{1j_{\tau(1)}} \dots C_{nj_{\tau(n)}}\\
    &=
    \sum_{\tau_f \in S_{n_f}}
    \sum_{\tau_b \in S_{n_b}}
    \sum_{\tau_{bf} \in S_n \setminus S_{n_b} \setminus S_{n_f}}
    \mathrm{sgn}(\tau_b) \mathrm{sgn}(\tau_f) \mathrm{sgn}(\tau_{bf})
    C_{1j_{\tau_b \tau_{bf}(1)}} \dots C_{n_b j_{\tau_b \tau_{bf}(n_b)}}
    C_{(n_b +1) j_{\tau_f \tau_{bf}(n_b + 1)}} \dots C_{n j_{\tau_f \tau_{bf}(n)}}\\
    &=
    \sum_{\tau_{bf} \in S_n \setminus S_{n_b} \setminus S_{n_f}}
    \mathrm{sgn}(\tau_{bf})
    \sum_{\tau_b \in S_{n_b}}
    \mathrm{sgn}(\tau_b)
    C_{1j_{\tau_b \tau_{bf}(1)}} \dots C_{n_b j_{\tau_b \tau_{bf}(n_b)}}
    \sum_{\tau_f \in S_{n_f}}
    \mathrm{sgn}(\tau_f)
    C_{(n_b+1)j_{\tau_f \tau_{bf}(n_b + 1)}} \dots C_{n j_{\tau_f \tau_{bf}(n)}}\\
    &=
    \sum_{\tau_{bf} \in S_n \setminus S_{n_b} \setminus S_{n_f}}
    \begin{vmatrix}
      C_{1j_{\tau_{bf}(1)}} & \dots & C_{1j_{\tau_{bf} (n_b)}}\\
      \vdots & \ddots & \vdots\\
      C_{n_b j_{\tau_{bf}(1)}} & \dots & C_{n_b j_{\tau_{bf}(n_b)}}\\
    \end{vmatrix}^+
    \begin{vmatrix}
      C_{(n_b+1) j_{\tau_{bf}(n_b+1)}} & \dots & C_{(n_b+1) j_{\tau_{bf}(n)}}\\
      \vdots & \ddots & \vdots\\
      C_{n j_{\tau_{bf}(n_b+1)}} & \dots & C_{n j_{\tau_{bf}(n)}}\\
    \end{vmatrix}^-\\
  \end{split}
\end{equation}
where $S_n$, $S_{n_b}$, and $S_{n_f}$ are sets of permutations of all indices,
first $n_b$ indices, and the remaining $n_f$ indices, respectively.
Then $S_n \setminus S_{n_b} \setminus S_{n_f}$ is the set of permutations
of the first $n_b$ columns with the rest.

In other words, $\tau_{bf}$ accounts for the different ways in which the columns of
the coefficient matrix can be split into two, one for $\tau_b$ and other for
$\tau_f$.
Since only the first $n_b$ rows are needed for the sum over $\tau_b$ and the
remaining rows for the sum over $\tau_f$, we will denote $C^b$ as the submatrix composed
of the first $n_b$ rows of $C$, and $C^f$ as the submatrix composed of the
remaining rows.
Note that $C^b$ has dimensions $n_b \times n$ and $C^f$ has dimensions
$n_f \times n$.
Then, we can rewrite the sum over $\tau_{bf}$ to be somewhat more transparent:
\begin{equation}
  \label{eq:permdet}
  \begin{split}
    \sum_{\tau \in S_n}
    \mathrm{sgn}(\tau)
    C_{1j_{\tau(1)}} \dots C_{nj_{\tau(n)}}
    &=
    \hspace{-3em}
    \sum_{
      \begin{smallmatrix}
        \{i^b_1 \dots i^b_{n_f}\} \subseteq \{j_1 \dots j_n\}\\
        \{i^f_1 \dots i^f_{n_b}\} = \{j_1 \dots j_n\} \setminus \{i^b_1 \dots i^b_{n_f}\}\\
      \end{smallmatrix}
    }
    \begin{vmatrix}
      C^b_{1 i^b_1} & \hspace{-0.6em}\dots & \hspace{-0.6em}C^b_{1 i^b_{n_b}}\\[-0.2em]
      \vdots & \hspace{-0.6em}\ddots & \hspace{-0.6em}\vdots\\[-0.6em]
      C^b_{n_b i^b_1} & \hspace{-0.6em}\dots & \hspace{-0.6em}C^b_{n_b i^b_{n_b}}\\
    \end{vmatrix}^+
    \begin{vmatrix}
      C^f_{1 i^f_1} &  \hspace{-0.6em}\dots &  \hspace{-0.4em}C^f_{1 i^f_{n_f}}\\[-0.2em]
      \vdots &  \hspace{-0.6em}\ddots &  \hspace{-0.4em}\vdots\\[-0.6em]
      C^f_{n_f i^f_1} &  \hspace{-0.6em}\dots &  \hspace{-0.4em}C^f_{n_f i^f_{n_f}}\\
    \end{vmatrix}^-
  \end{split}
\end{equation}
% I don't think we can actually do this
% Alternatively, $\tau_{bf}$ can swap rows instead of columns
% \begin{equation}
%   \label{eq:permdet2}
%   \begin{split}
%     \sum_{\tau \in S_n}
%     \mathrm{sgn}(\tau)
%     C_{1j_{\tau(1)}} \dots C_{nj_{\tau(n)}}
%     &=
%     \hspace{-2.5em}
%     \sum_{
%       \begin{smallmatrix}
%         \{i^b_1 \dots i^b_{n_f}\} \subseteq \{1 \dots n\}\\
%         \{i^f_1 \dots i^f_{n_b}\} = \{1 \dots n\} \setminus \{i^b_1 \dots i^b_{n_f}\}\\
%       \end{smallmatrix}
%     }
%     \begin{vmatrix}
%       C_{i^b_1 j_1} & \hspace{-0.6em}\dots & \hspace{-0.6em}C_{i^b_1 j_{n_b}}\\
%       \vdots & \hspace{-0.6em}\ddots & \hspace{-0.6em}\vdots\\[-0.6em]
%       C_{i^b_{n_b} j_1} & \hspace{-0.6em}\dots & \hspace{-0.6em}C_{i^b_{n_b} j_{n_b}}\\
%     \end{vmatrix}^+
%     \begin{vmatrix}
%       C_{i^f_1 j_{n_b+1}} & \hspace{-0.6em}\dots & \hspace{-0.4em}C_{i^f_1 j_n}\\
%       \vdots & \hspace{-0.6em}\ddots & \hspace{-0.4em}\vdots\\[-0.6em]
%       C_{i^f_{n_f} j_{n_b+1}} & \hspace{-0.6em}\dots & \hspace{-0.4em}C_{i^f_{n_f} j_n}\
%     \end{vmatrix}^-
%   \end{split}
% \end{equation}

If the given operators do not commute or anticommute with one another,
then an explicit sum through every permutation might be necessary, along with
the signature of each permutation, unless there is a specialized structure that
simplifies this sum.

\subsection{CC}
\label{sec:appendix_cc}
\begin{equation}
  \label{eq:cc_deriv}
  \begin{split}
    \ket{\Psi_{\mathrm{CC}}} &=
    \exp \left(
      \sum_{\mathbf{i} \mathbf{a}} t_{\mathbf{i}}^{\mathbf{a}} \hat{E}_{\mathbf{i}}^{\mathbf{a}}
    \right)
    \ket{\Phi_{\mathrm{HF}}}\\
    &= \sum_{n=0}^\infty \frac{1}{n!}
    \left(
      \sum_{\mathbf{i} \mathbf{a}} t_{\mathbf{i}}^{\mathbf{a}} \hat{E}_{\mathbf{i}}^{\mathbf{a}}
    \right)^{n}
    \ket{\Phi_{\mathrm{HF}}}\\
    &= \sum_{n=0}^\infty \frac{1}{n!}
    \prod_{k=1}^n
    \sum_{\mathbf{i} \mathbf{a}} t_{\mathbf{i}}^{\mathbf{a}} \hat{E}_{\mathbf{i}}^{\mathbf{a}}
    \ket{\Phi_{\mathrm{HF}}}
  \end{split}
\end{equation}
For each $n$, we can treat
$\prod_{k=1}^n \sum_{\mathbf{i} \mathbf{a}} t_{\mathbf{i}}^{\mathbf{a}} \hat{E}_{\mathbf{i}}^{\mathbf{a}}$
with the approach from Equation~\ref{eq:approach}.
Then, the operators, $\{\hat{Q}_{j_1} \dots \hat{Q}_{j_n}\}$, are excitation
operators, $\{\hat{E}_{\mathbf{i}_1}^{\mathbf{a}_1} \dots \hat{E}_{\mathbf{i}_n}^{\mathbf{a}_n}\}$,
and $n$ controls the number of operators used in the second summation of
Equation~\ref{eq:approach}.
Since they all commute with one another, $\mathrm{sgn}(\tau) = 1$ which means that
the final sum of Equation~\ref{eq:approach} will be a permanent
(Equation~\ref{eq:FANCI_cc_f}).

\begin{equation}
  \begin{split}
    \prod_{k=1}^n \sum_{\mathbf{i} \mathbf{a}} t_{\mathbf{i}}^{\mathbf{a}} \hat{E}_{\mathbf{i}}^{\mathbf{a}}
    &= \sum_{\mathbf{m}}
    \sum_{
      \begin{smallmatrix}
        \{\hat{E}_{\mathbf{i}_1}^{\mathbf{a}_1} \dots \hat{E}_{\mathbf{i}_n}^{\mathbf{a}_n}\}
        \subseteq \tilde{S}_{\hat{E}}\\
        \mathrm{sgn} \prod_{k=1}^n \hat{E}_{\mathbf{i}_k}^{\mathbf{a}_k} \ket{\Phi_{\mathrm{HF}}} = \ket{\mathbf{m}}
      \end{smallmatrix}
    }
    \mathrm{sgn}\big( \sigma(\hat{E}_{\mathbf{i}_1}^{\mathbf{a}_1} \dots \hat{E}_{\mathbf{i}_n}^{\mathbf{a}_n})\big)
    \sum_{\tau \in S_n}
    t_{\mathbf{i}_{\tau(1)}}^{\mathbf{a}_{\tau(1)}} \dots t_{\mathbf{i}_{\tau(n)}}^{\mathbf{a}_{\tau(n)}}
    \ket{\mathbf{m}}\\
    &= \sum_{\mathbf{m}}
    \sum_{
      \begin{smallmatrix}
        \{\hat{E}_{\mathbf{i}_1}^{\mathbf{a}_1} \dots \hat{E}_{\mathbf{i}_n}^{\mathbf{a}_n}\}
        \subseteq \tilde{S}_{\hat{E}}\\
        \mathrm{sgn} \prod_{k=1}^n \hat{E}_{\mathbf{i}_k}^{\mathbf{a}_k} \ket{\Phi_{\mathrm{HF}}} = \ket{\mathbf{m}}
      \end{smallmatrix}
    }
    \mathrm{sgn}\big( \sigma(\hat{E}_{\mathbf{i}_1}^{\mathbf{a}_1} \dots \hat{E}_{\mathbf{i}_n}^{\mathbf{a}_n})\big)
    n! \prod_{k=1}^n t_{\mathbf{i}_k}^{\mathbf{a}_k}
    \ket{\mathbf{m}}
  \end{split}
\end{equation}
where $\tilde{S}_{\hat{E}}$ is the set of all excitation operators used in the
wavefunction.
Substituting into the Equation~\ref{eq:cc_deriv}, we get
\begin{equation}
  \label{eq:cc_deriv2}
  \begin{split}
    \ket{\Psi_{\mathrm{CC}}}
    &= \sum_{n=0}^\infty \frac{1}{n!}
    \sum_{\mathbf{m}}
    \sum_{
      \begin{smallmatrix}
        \{\hat{E}_{\mathbf{i}_1}^{\mathbf{a}_1} \dots \hat{E}_{\mathbf{i}_n}^{\mathbf{a}_n}\}
        \subseteq \tilde{S}_{\hat{E}}\\
        \mathrm{sgn} \prod_{k=1}^n \hat{E}_{\mathbf{i}_k}^{\mathbf{a}_k} \ket{\Phi_{\mathrm{HF}}} = \ket{\mathbf{m}}
      \end{smallmatrix}
    }
    \mathrm{sgn}\big( \sigma(\hat{E}_{\mathbf{i}_1}^{\mathbf{a}_1} \dots \hat{E}_{\mathbf{i}_n}^{\mathbf{a}_n})\big)
    n! \prod_{k=1}^n t_{\mathbf{i}_k}^{\mathbf{a}_k}
    \ket{\mathbf{m}}\\
    &=
    \sum_{\mathbf{m}}
    \sum_{n=0}^\infty
    \sum_{
      \begin{smallmatrix}
        \{\hat{E}_{\mathbf{i}_1}^{\mathbf{a}_1} \dots \hat{E}_{\mathbf{i}_n}^{\mathbf{a}_n}\}
        \subseteq \tilde{S}_{\hat{E}}\\
        \mathrm{sgn} \prod_{k=1}^n \hat{E}_{\mathbf{i}_k}^{\mathbf{a}_k} \ket{\Phi_{\mathrm{HF}}} = \ket{\mathbf{m}}
      \end{smallmatrix}
    }
    \mathrm{sgn}\big( \sigma(\hat{E}_{\mathbf{i}_1}^{\mathbf{a}_1} \dots \hat{E}_{\mathbf{i}_n}^{\mathbf{a}_n})\big)
    \prod_{k=1}^n t_{\mathbf{i}_k}^{\mathbf{a}_k}
    \ket{\mathbf{m}}\\
    &=
    \sum_{\mathbf{m}}
    \sum_{
      \begin{smallmatrix}
        \{\hat{E}_{\mathbf{i}_1}^{\mathbf{a}_1} \dots \hat{E}_{\mathbf{i}_n}^{\mathbf{a}_n} | n \in \mathbb{N}\}
        \subseteq \tilde{S}_{\hat{E}}\\
        \mathrm{sgn} \prod_{k=1}^n \hat{E}_{\mathbf{i}_k}^{\mathbf{a}_k} \ket{\Phi_{\mathrm{HF}}} = \ket{\mathbf{m}}
      \end{smallmatrix}
    }
    \mathrm{sgn}\big( \sigma(\hat{E}_{\mathbf{i}_1}^{\mathbf{a}_1} \dots \hat{E}_{\mathbf{i}_n}^{\mathbf{a}_n})\big)
    \prod_{k=1}^n t_{\mathbf{i}_k}^{\mathbf{a}_k}
    \ket{\mathbf{m}}\\
  \end{split}
\end{equation}
where the second sum is the sum over all possible combinations of excitations
that would produce the given Slater determinant from the reference.
Though it is unnecessary, $n \in \mathbb{N}$ was included to specify that the
subset is no longer constrained to be of a certain size by an external variable,
$n$.
In the rest of the article, this notation will be dropped.

\subsection{CC with Creation Operators}
\label{sec:appendix_cc_creation}
Let $\mathbf{b}_i$ be a set of an even number of orbitals, $\mathbf{f}_i$ be a set
of an odd number of orbitals, $\hat{A}^\dagger_{\mathbf{b}_i}$ and
$\hat{A}^\dagger_{\mathbf{f}_i}$ be the creation operators that create these
orbitals, and $C_{\mathbf{b}_i}$ and $C_{\mathbf{f}_i}$ are the
coefficient for the associated creation operators.
Then, the CC wavefunction using creation operators will be
\begin{equation}
  \ket{\Psi_{\mathrm{CC}}} = \exp
  \left(
    \sum_{\mathbf{b}_i} C_{\mathbf{b}_i} \hat{A}^\dagger_{\mathbf{b}_i} +
    \sum_{\mathbf{f}_i} C_{\mathbf{f}_i} \hat{A}^\dagger_{\mathbf{f}_i}
  \right)
  \ket{0}
\end{equation}

Taking the same approach as Equation~\ref{eq:approach}, we have
operators, $\{\hat{Q}_{j_1} \dots \hat{Q}_{j_n}\}$, that are creation
operators,
$\{\hat{A}_{\mathbf{b}_1} \dots \hat{A}_{\mathbf{b}_{n_b}}
\hat{A}_{\mathbf{f}_1} \dots \hat{A}_{\mathbf{f}_{n_f}}\}$, and the sum over
the permutation is given by Equation~\ref{eq:permdet}.
Following Equation~\ref{eq:approach} and \ref{eq:permdet},
\begin{equation}
  \begin{split}
    \hspace{-3em}
    \ket{\Psi}
    &=
    \sum_{\mathbf{m}}
    \sum_{n=0}^\infty
    \frac{1}{n!}
    \hspace{-1.5em}
    \sum_{
      \begin{smallmatrix}
        \\[0.2em]
        \{\mathbf{b}_1 \dots \mathbf{b}_{n_b} \} \subseteq \tilde{S}_{\mathbf{b}}, \;
        \{\mathbf{f}_1 \dots \mathbf{f}_{n_f} \} \subseteq \tilde{S}_{\mathbf{f}}\\
        \mathrm{sgn}
        \hat{A}^\dagger_{\mathbf{b}_1} \dots \hat{A}^\dagger_{\mathbf{b}_{n_b}}
        \hat{A}^\dagger_{\mathbf{f}_1} \dots \hat{A}^\dagger_{\mathbf{f}_{n_f}}
        \ket{0}
        = \ket{\mathbf{m}}
      \end{smallmatrix}
    }
    \hspace{-4em}
    \mathrm{sgn}
    \left(
      \sigma(
      \hat{A}^\dagger_{\mathbf{b}_1} \dots \hat{A}^\dagger_{\mathbf{b}_{n_b}}
      \hat{A}^\dagger_{\mathbf{f}_1} \dots \hat{A}^\dagger_{\mathbf{f}_{n_f}}
      )
    \right)
    \hspace{-2.5em}
    \sum_{
      \begin{smallmatrix}
        \{i^b_1 \dots i^b_{n_b}\} \subseteq \{1 \dots n\}\\
        \{i^f_1 \dots i^f_{n_f}\} = \{1 \dots n\} \setminus \{i^b_1 \dots i^b_{n_b}\}\\
      \end{smallmatrix}
    }
    \begin{vmatrix}
      C_{\mathbf{b}_1} & \hspace{-0.6em}\dots & \hspace{-0.6em}C_{\mathbf{b}_{n_b}}\\[-0.2em]
      \vdots & \hspace{-0.6em}\ddots & \hspace{-0.6em}\vdots\\[-0.6em]
      C_{\mathbf{b}_1} & \hspace{-0.6em}\dots & \hspace{-0.6em}C_{\mathbf{b}_{n_b}}\\
    \end{vmatrix}^+
    \begin{vmatrix}
      C_{\mathbf{f}_1} & \hspace{-0.6em}\dots & \hspace{-0.4em}C_{\mathbf{f}_{n_f}}\\[-0.2em]
      \vdots & \hspace{-0.6em}\ddots & \hspace{-0.4em}\vdots\\[-0.6em]
      C_{\mathbf{f}_1} & \hspace{-0.6em}\dots & \hspace{-0.4em}C_{\mathbf{f}_{n_f}}\\
    \end{vmatrix}^-
    \hspace{-0.8em}
    \ket{\mathbf{m}}\\
    &=
    \sum_{\mathbf{m}}
    \sum_{
      \begin{smallmatrix}
        \{\mathbf{b}_1 \dots \mathbf{b}_{n_b} \} \subseteq \tilde{S}_{\mathbf{b}}, \;
        \{\mathbf{f}_1 \dots \mathbf{f}_{n_f} \} \subseteq \tilde{S}_{\mathbf{f}}\\
        \mathrm{sgn}
        \hat{A}^\dagger_{\mathbf{b}_1} \dots \hat{A}^\dagger_{\mathbf{b}_{n_b}}
        \hat{A}^\dagger_{\mathbf{f}_1} \dots \hat{A}^\dagger_{\mathbf{f}_{n_f}}
        \ket{0}
        = \ket{\mathbf{m}}
      \end{smallmatrix}
    }
    \hspace{-4em}
    \mathrm{sgn}
    \left(
      \sigma(
      \hat{A}^\dagger_{\mathbf{b}_1} \dots \hat{A}^\dagger_{\mathbf{b}_{n_b}}
      \hat{A}^\dagger_{\mathbf{f}_1} \dots \hat{A}^\dagger_{\mathbf{f}_{n_f}}
      )
    \right)
    \frac{1}{n!}
    \binom{n}{n_b}
    \begin{vmatrix}
      C_{\mathbf{b}_1} & \hspace{-0.6em}\dots & \hspace{-0.6em}C_{\mathbf{b}_{n_b}}\\[-0.2em]
      \vdots & \hspace{-0.6em}\ddots & \hspace{-0.6em}\vdots\\[-0.6em]
      C_{\mathbf{b}_1} & \hspace{-0.6em}\dots & \hspace{-0.6em}C_{\mathbf{b}_{n_b}}\\
    \end{vmatrix}^+
    \begin{vmatrix}
      C_{\mathbf{f}_1} & \hspace{-0.6em}\dots & \hspace{-0.4em}C_{\mathbf{f}_{n_f}}\\[-0.2em]
      \vdots & \hspace{-0.6em}\ddots & \hspace{-0.4em}\vdots\\[-0.6em]
      C_{\mathbf{f}_1} & \hspace{-0.6em}\dots & \hspace{-0.4em}C_{\mathbf{f}_{n_f}}\\
    \end{vmatrix}^-
    \hspace{-0.8em}
    \ket{\mathbf{m}}\\
    &=
    \sum_{\mathbf{m}}
    \sum_{
      \begin{smallmatrix}
        \{\mathbf{b}_1 \dots \mathbf{b}_{n_b} \} \subseteq \tilde{S}_{\mathbf{b}}, \;
        \{\mathbf{f}_1 \dots \mathbf{f}_{n_f} \} \subseteq \tilde{S}_{\mathbf{f}}\\
        \mathrm{sgn}
        \hat{A}^\dagger_{\mathbf{b}_1} \dots \hat{A}^\dagger_{\mathbf{b}_{n_b}}
        \hat{A}^\dagger_{\mathbf{f}_1} \dots \hat{A}^\dagger_{\mathbf{f}_{n_f}}
        \ket{0}
        = \ket{\mathbf{m}}
      \end{smallmatrix}
    }
    \hspace{-4em}
    \mathrm{sgn}
    \left(
      \sigma(
      \hat{A}^\dagger_{\mathbf{b}_1} \dots \hat{A}^\dagger_{\mathbf{b}_{n_b}}
      \hat{A}^\dagger_{\mathbf{f}_1} \dots \hat{A}^\dagger_{\mathbf{f}_{n_f}}
      )
    \right)
    \frac{1}{n_b! n_f!}
    \begin{vmatrix}
      C_{\mathbf{b}_1} & \hspace{-0.6em}\dots & \hspace{-0.6em}C_{\mathbf{b}_{n_b}}\\[-0.2em]
      \vdots & \hspace{-0.6em}\ddots & \hspace{-0.6em}\vdots\\[-0.6em]
      C_{\mathbf{b}_1} & \hspace{-0.6em}\dots & \hspace{-0.6em}C_{\mathbf{b}_{n_b}}\\
    \end{vmatrix}^+
    \begin{vmatrix}
      C_{\mathbf{f}_1} & \hspace{-0.6em}\dots & \hspace{-0.4em}C_{\mathbf{f}_{n_f}}\\[-0.2em]
      \vdots & \hspace{-0.6em}\ddots & \hspace{-0.4em}\vdots\\[-0.6em]
      C_{\mathbf{f}_1} & \hspace{-0.6em}\dots & \hspace{-0.4em}C_{\mathbf{f}_{n_f}}\\
    \end{vmatrix}^-
    \hspace{-0.8em}
    \ket{\mathbf{m}}\\
  \end{split}
\end{equation}
Note that the $n \in \mathbb{N}$ was omitted and it is implied that the size of
the subset is not constrained by an external variable.

Since the determinant is zero when any two rows are equivalent, $n_f \leq 1$.
In contrast, we can take the permanent of a matrix with repeating rows:
\begin{equation}
  \begin{split}
    \begin{vmatrix}
      a_{1} & \dots & a_{n}\\
      \vdots & \ddots & \vdots\\
      a_{1} & \dots & a_{n}\\
    \end{vmatrix}^+
    &= \sum_{\sigma \in S_n} \prod_{i=1} a_{i}\\
    &= n! \prod_{i=1} a_{i}
  \end{split}
\end{equation}
Therefore, we get
\begin{equation}
  \begin{split}
    \hspace{-3em}
    \ket{\Psi}
    &=
    \sum_{\mathbf{m}}
    \begin{pmatrix}
      \mathlarger{\sum}\limits_{
        \begin{smallmatrix}
          \\[0.2em]
          \{\mathbf{b}_1 \dots \mathbf{b}_{n_b} \} \subseteq \tilde{S}_{\mathbf{b}}, \;
          \mathbf{f} \in \tilde{S}_{\mathbf{f}}\\
          \mathrm{sgn}
          \hat{A}^\dagger_{\mathbf{b}_1} \dots \hat{A}^\dagger_{\mathbf{b}_{n_b}}
          \hat{A}^\dagger_{\mathbf{f}}
          \ket{0}
          = \ket{\mathbf{m}}
        \end{smallmatrix}
      }
      \hspace{-3.5em}
      \mathrm{sgn}
      \left(
        \sigma(
        \hat{A}^\dagger_{\mathbf{b}_1} \dots \hat{A}^\dagger_{\mathbf{b}_{n_b}}
        \hat{A}^\dagger_{\mathbf{f}}
        )
      \right)
      \left(
        \prod\limits_{i=1}^{n_b} C_{\mathbf{b}_i}
      \right)
      C_{\mathbf{f}}
      &\\
      & \hspace{-15em}
      +
      \mathlarger{\sum}\limits_{
        \begin{smallmatrix}
          \\[0.2em]
          \{\mathbf{b}_1 \dots \mathbf{b}_{n_b} \} \subseteq \tilde{S}_{\mathbf{b}}\\
          \mathrm{sgn}
          \hat{A}^\dagger_{\mathbf{b}_1} \dots \hat{A}^\dagger_{\mathbf{b}_{n_b}}
          \ket{0}
          = \ket{\mathbf{m}}
        \end{smallmatrix}
      }
      \hspace{-3em}
      \mathrm{sgn}
      \left(
        \sigma(
        \hat{A}^\dagger_{\mathbf{b}_1} \dots \hat{A}^\dagger_{\mathbf{b}_{n_b}}
        )
      \right)
      \prod\limits_{i=1}^{n_b} C_{\mathbf{b}_i}
    \end{pmatrix}
    \ket{\mathbf{m}}\\
  \end{split}
\end{equation}
Since the sum over $\{\mathbf{b}_1 \dots \mathbf{b}_{n_b} \mathbf{f}\}$
only occurs when $\mathbf{m}$ has an odd number of electrons and the sum over
$\{\mathbf{b}_1 \dots \mathbf{b}_{n_b}\}$ only occurs when $\mathbf{m}$ has an
even number of electrons, we can separate these two cases if the desired number
of electrons in the system, $N$, is either even or odd.
\begin{equation}
  \hspace{-3em}
  \ket{\Psi}
  =
  \begin{cases}
    \mathlarger{\sum}\limits_{\mathbf{m}}
    \mathlarger{\sum}\limits_{
      \begin{smallmatrix}
        \\[0.2em]
        \{\mathbf{b}_1 \dots \mathbf{b}_{n_b} \} \subseteq \tilde{S}_{\mathbf{b}}, \;
        \mathbf{f} \in \tilde{S}_{\mathbf{f}}\\
        \mathrm{sgn}
        \hat{A}^\dagger_{\mathbf{b}_1} \dots \hat{A}^\dagger_{\mathbf{b}_{n_b}}
        \hat{A}^\dagger_{\mathbf{f}}
        \ket{0}
        = \ket{\mathbf{m}}
      \end{smallmatrix}
    }
    \hspace{-3.5em}
    \mathrm{sgn}
    \left(
      \sigma(
      \hat{A}^\dagger_{\mathbf{b}_1} \dots \hat{A}^\dagger_{\mathbf{b}_{n_b}}
      \hat{A}^\dagger_{\mathbf{f}}
      )
    \right)
    \left(
      \prod\limits_{i=1}^{n_b} C_{\mathbf{b}_i}
    \right)
    C_{\mathbf{f}}
    \ket{\mathbf{m}}
    & \mbox{if $N$ is odd}\\[1.4em]
    \mathlarger{\sum}\limits_{\mathbf{m}}
    \mathlarger{\sum}\limits_{
      \begin{smallmatrix}
        \\[0.2em]
        \{\mathbf{b}_1 \dots \mathbf{b}_{n_b} \} \subseteq \tilde{S}_{\mathbf{b}}\\
        \mathrm{sgn}
        \hat{A}^\dagger_{\mathbf{b}_1} \dots \hat{A}^\dagger_{\mathbf{b}_{n_b}}
        \ket{0}
        = \ket{\mathbf{m}}
      \end{smallmatrix}
    }
    \hspace{-3em}
    \mathrm{sgn}
    \left(
      \sigma(
      \hat{A}^\dagger_{\mathbf{b}_1} \dots \hat{A}^\dagger_{\mathbf{b}_{n_b}}
      )
    \right)
    \prod\limits_{i=1}^{n_b} C_{\mathbf{b}_i}
    \ket{\mathbf{m}}
    & \mbox{if $N$ is even}\\
  \end{cases}
\end{equation}

\subsection{Generalized Quasiparticle}
\label{sec:appendix_quasiparticle}
Just as in Section~\ref{sec:appendix_cc_creation}, let $\mathbf{b}_i$
be a set of an even number of orbitals, $\mathbf{f}_i$ be a set
of an odd number of orbitals, $\hat{A}^\dagger_{\mathbf{b}_i}$ and
$\hat{A}^\dagger_{\mathbf{f}_i}$ be the creation operators that create the
associated orbitals.
We can construct a quasiparticle as a linear combination of these creation
operators.
The desired wavefunction is a product of these quasiparticles
\begin{equation}
  \ket{\Psi} = \prod_{p=1}^n
  \left(
    \sum_{\mathbf{b}_i} C_{p;\mathbf{b}_i} \hat{A}^\dagger_{\mathbf{b}_i} +
    \sum_{\mathbf{f}_i} C_{p;\mathbf{f}_i} \hat{A}^\dagger_{\mathbf{f}_i}
  \right)
  \ket{0}
\end{equation}
where $C_{p;\mathbf{b}_i}$ and $C_{p;\mathbf{f}_i}$ are coefficients of the
creation operators $\hat{A}^\dagger_{\mathbf{b}_i}$ and
$\hat{A}^\dagger_{\mathbf{f}_i}$ in the construction of the
$p$\textsuperscript{th} quasiparticle.

Taking the same approach as Equation~\ref{eq:approach}, we have
operators, $\{\hat{Q}_{j_1} \dots \hat{Q}_{j_n}\}$, that are creation
operators,
$\{\hat{A}_{\mathbf{b}_1} \dots \hat{A}_{\mathbf{b}_{n_b}}
\hat{A}_{\mathbf{f}_1} \dots \hat{A}_{\mathbf{f}_{n_f}}\}$, and the sum over
the permutation is given by Equation~\ref{eq:permdet}.
\begin{equation}
  \begin{split}
    \hspace{-4em}
    \ket{\Psi}
    &=
    \sum_{\mathbf{m}}
    \hspace{-1em}
    \mathlarger{\sum}_{
      \begin{smallmatrix}
        \\[0.2em]
        \{\mathbf{b}_1 \dots \mathbf{b}_{n_b} \} \subseteq \tilde{S}_{\mathbf{b}}, \;
        \{\mathbf{f}_1 \dots \mathbf{f}_{n_f} \} \subseteq \tilde{S}_{\mathbf{f}}\\
        \mathrm{sgn}
        \hat{A}^\dagger_{\mathbf{b}_1} \dots \hat{A}^\dagger_{\mathbf{b}_{n_b}}
        \hat{A}^\dagger_{\mathbf{f}_1} \dots \hat{A}^\dagger_{\mathbf{f}_{n_f}}
        \ket{0}
        = \ket{\mathbf{m}}
      \end{smallmatrix}
    }
    \hspace{-4.5em}
    \mathrm{sgn}
    \left(
      \sigma(
      \hat{A}^\dagger_{\mathbf{b}_1} \dots \hat{A}^\dagger_{\mathbf{b}_{n_b}}
      \hat{A}^\dagger_{\mathbf{f}_1} \dots \hat{A}^\dagger_{\mathbf{f}_{n_f}}
      )
    \right)
    \hspace{-5em}
    \mathlarger{\sum}_{
      \begin{smallmatrix}
        \{i^b_1 \dots i^b_{n_b}\} \subseteq
        \{\mathbf{b}_1 \dots \mathbf{b}_{n_b} \mathbf{f}_1 \dots \mathbf{f}_{n_f} \}\\
        \{i^f_1 \dots i^f_{n_f}\} =
        \{\mathbf{b}_1 \dots \mathbf{b}_{n_b} \mathbf{f}_1 \dots \mathbf{f}_{n_f} \}
        \setminus \{i^b_1 \dots i^b_{n_b}\}\\
      \end{smallmatrix}
    }
    \begin{vmatrix}
      C^b_{1 i^b_1} & \hspace{-0.6em}\dots & \hspace{-0.6em}C^b_{1 i^b_{n_b}}\\[-0.2em]
      \vdots & \hspace{-0.6em}\ddots & \hspace{-0.6em}\vdots\\[-0.6em]
      C^b_{n_b i^b_1} & \hspace{-0.6em}\dots & \hspace{-0.6em}C^b_{n_b i^b_{n_b}}\\
    \end{vmatrix}^+
    \begin{vmatrix}
      C^f_{1 i^f_1} & \hspace{-0.6em}\dots & \hspace{-0.4em}C^f_{1 i^f_{n_f}}\\[-0.2em]
      \vdots & \hspace{-0.6em}\ddots & \hspace{-0.4em}\vdots\\[-0.6em]
      C^f_{n_f i^f_1} & \hspace{-0.6em}\dots & \hspace{-0.4em}C^f_{n_f i^f_{n_f}}\\
    \end{vmatrix}^-
    \hspace{-0.8em}
    \ket{\mathbf{m}}\\
  \end{split}
\end{equation}
where $C^b$ is the submatrix of $C$ composed of the first $n_b$ rows and $C^f$
is composed of the remaining rows.
It was assumed that the first $n_b$ columns belong to the even-electron operators and
the remaining columns to the odd-electron operators.

\end{document}